\documentclass[a4paper,fleqn,usenatbib]{mnras}

\usepackage[T1]{fontenc}
\usepackage{ae,aecompl}

\usepackage{graphicx}	% Including figure files
\usepackage{savesym}
\usepackage{amsmath}	% Advanced maths commands
\usepackage{amssymb}	% Extra maths symbols
\usepackage{amstext}
\savesymbol{iint}
\usepackage{txfonts}
\restoresymbol{TXF}{iint}

\usepackage{pdflscape} % Paquete para girar la pagina para las tablas
\usepackage{longtable} % para tablas largas

\newcommand{\elecd}{$n_{\rm e}$}
\newcommand{\elect}{$T_{\rm e}$}
\newcommand{\tf}{$t^{2}$}
\newcommand{\hb}{H$\beta$}
\newcommand{\ha}{H$\alpha$}

\newcommand{\foiii}{[O\thinspace{\sc iii}]}

\newcommand{\foi}{[O\thinspace{\sc i}]}
\newcommand{\foii}{[O\thinspace{\sc ii}]}
\newcommand{\fsii}{[S\thinspace{\sc ii}]}
\newcommand{\fsiii}{[S\thinspace{\sc iii}]}
\newcommand{\fni}{[N\thinspace{\sc i}]}
\newcommand{\fnii}{[N\thinspace{\sc ii}]}
\newcommand{\fariv}{[Ar\thinspace{\sc iv}]}
\newcommand{\fcliii}{[Cl\thinspace{\sc iii}]}

\newcommand{\fneiii}{[Ne\thinspace{\sc iii}]}
\newcommand{\ffei}{[Fe\thinspace{\sc i}]}
\newcommand{\ffeii}{[Fe\thinspace{\sc ii}]}
\newcommand{\ffeiii}{[Fe\thinspace{\sc iii}]}

\newcommand{\fniqii}{[Ni\thinspace{\sc ii}]}
\newcommand{\fniqiii}{[Ni\thinspace{\sc iii}]}

\newcommand{\nitroi}{N\thinspace{\sc i}}
\newcommand{\nii}{N\thinspace{\sc ii}}
\newcommand{\niii}{N\thinspace{\sc iii}}

\newcommand{\oi}{O\thinspace{\sc i}}
\newcommand{\oii}{O\thinspace{\sc ii}}

\newcommand{\cii}{C\thinspace{\sc ii}}
\newcommand{\ciii}{C\thinspace{\sc iii}}

\newcommand{\fariii}{[Ar\thinspace{\sc iii}]}

\newcommand{\hi}{H\,{\sc i}}
\newcommand{\hii}{H\thinspace{\sc ii}}

\newcommand{\hei}{He\thinspace{\sc i}}
\newcommand{\heii}{He\thinspace{\sc ii}}

\newcommand{\ts}{\emph{$t^2$}}

\newcommand\ionic[2]{${\rm #1^{#2}}$}           % ej. \ionic{O}{+} --> O^+

\newcommand{\cmc}{{\rm cm$^{-3}$}}
\newcommand{\kms}{{\rm km~s$^{-1}$}}

\title[Chemical composition of Galactic ring nebulae]{The chemical composition of Galactic ring nebulae around massive stars}

\author[C. Esteban et al.]{
C. Esteban,$^{1,2}$\thanks{E-mail: cel@iac.es}
A. Mesa-Delgado,$^{3}$
C. Morisset$^{4}$
and J. Garc{\'{\i}}a-Rojas$^{1,2}$
\\
$^{1}$Instituto de Astrof\'isica de Canarias, E-38200 La Laguna, Tenerife, Spain\\
$^{2}$Departamento de Astrof\'isica, Universidad de La Laguna, E-38206, La Laguna, Tenerife, Spain\\
$^{3}$Instituto de Astrof\'isica, Facultad de F\'isica, Pontificia Universidad Cat\'olica de Chile, Av.~Vicu\~na Mackenna 4860,\\ 782-0436 Macul, Santiago, Chile\\
$^{4}$Instituto de Astronom\'ia, Universidad Nacional Aut\'onoma de M\'exico, Apdo. Postal 70264, 04510 M\'exico D.F., Mexico
}

\date{Accepted XXX. Received YYY; in original form ZZZ}

\pubyear{2016}

\begin{document}
\label{firstpage}
\pagerange{\pageref{firstpage}--\pageref{lastpage}}
\maketitle

\begin{abstract}
We present deep spectra of ring nebulae associated with Wolf-Rayet (WR) and O-type stars: NGC~6888, G2.4+1.4, RCW~58, S~308, NGC~7635 and RCW~52. The data have been taken with the 10m 
Gran Telescopio Canarias and the 6.5m Clay Telescope. We extract spectra of several apertures in some of the objects. We derive C$^{2+}$ and O$^{2+}$ abundances from faint recombination lines in NGC~6888 and NGC~7635, permitting to derive their C/H and C/O ratios and estimate the abundance discrepancy factor (ADF) of O$^{2+}$.  The ADFs are larger than the typical ones of normal {\hii} regions but similar to those found in the ionised gas of star-forming dwarf galaxies. We find that chemical abundances are rather homogeneous in the nebulae where we have spectra of several apertures: NGC~6888, NGC~7635 and G2.4+1.4. We obtain very high values of electron temperature in a peripheral zone of NGC~6888, finding that shock excitation can reproduce its spectral properties. We find that all the objects associated with WR stars show N enrichment. Some of them also show He enrichment and O deficiency as well as a lower Ne/O than expected, this may indicate the strong action of the ON and NeNa cycles. We have compared the chemical composition of NGC~6888, G2.4+1.4, RCW~58 and S~308 with the nucleosynthesis predicted by stellar evolution models of massive stars. We find that non-rotational models of stars of initial masses between 25 and 40 $M_{\odot}$ seem to reproduce the observed abundance ratios of most of the nebulae. 
\end{abstract}


%%%%%%%%%%%%%%%%%%%%%%%%%%%%%%%%%%%%%%%%%%%%%%%%%%

%%%%%%%%%%%%%%%%% BODY OF PAPER %%%%%%%%%%%%%%%%%%

%%%%%%%%%%%%%%%%%%%%%%%%%%%%%%%%%%%%%%%%%%%%%%%%%%%%%%%%%%%%%%%%%%%%%%%%%%%%%%%%%%%%%%%%%%%%%%%%%
\section{Introduction}
\label{sec:intro}
%%%%%%%%%%%%%%%%%%%%%%%%%%%%%%%%%%%%%%%%%%%%%%%%%%%%%%%%%%%%%%%%%%%%%%%%%%%%%%%%%%%%%%%%%%%%%%%%%

Ring nebulae are interstellar bubbles of ionised gas that have swept-up the surrounding circumstellar medium after the mass loss episodes experienced by their massive progenitors. They are 
rather common around Wolf-Rayet (WR) stars and luminous blue variables (LBVs) but are rarely seen around main-sequence O stars. This may depend on the density of the circumstellar material on which they develop \citep[e.g.][]{nazeetal01}. There have been several spectroscopical works devoted to study the chemical composition of ring nebulae around WR stars \citep[e.g.][]{kwitter84, estebanetal92, stocketal11} and LBVs \citep[e.g.][]{smithetal97, smithetal98}. These works have found that some ring nebulae show clear chemical enrichment patterns, in general they are He and N enriched and 
some of them show some O deficiency, indicating that they are composed by material ejected by the central star. Therefore, a detailed analysis of the chemical composition of stellar ejecta in ring nebulae would allow us to explore the nucleosynthesis of massive stars, constrain stellar evolution models and the evolutive scenario of the stellar progenitors. \citet{estebanvilchez92}, \citet{estebanetal92} and \citet{mesadelgadoetal14} compared the abundance pattern of Galactic ejecta ring nebulae with the surface abundances predicted by the evolutionary models of massive stars. These authors found that the He, N and O abundances of the nebulae are consistent with the expected nucleosynthesis for stars with initial masses between 25 and 40 M$_\odot$ at a given moment of 
the red supergiant (RSG) phase, prior to the onset of the WR stage. A similar result was obtained by \citet{garnettchu94} for the ejecta nebula associated to the WR star Br~13 in the 
Large Magellanic Cloud. 

%%%%%%%%%%%%%%% 
  \begin{figure}
   \centering
   \includegraphics[width=85mm]{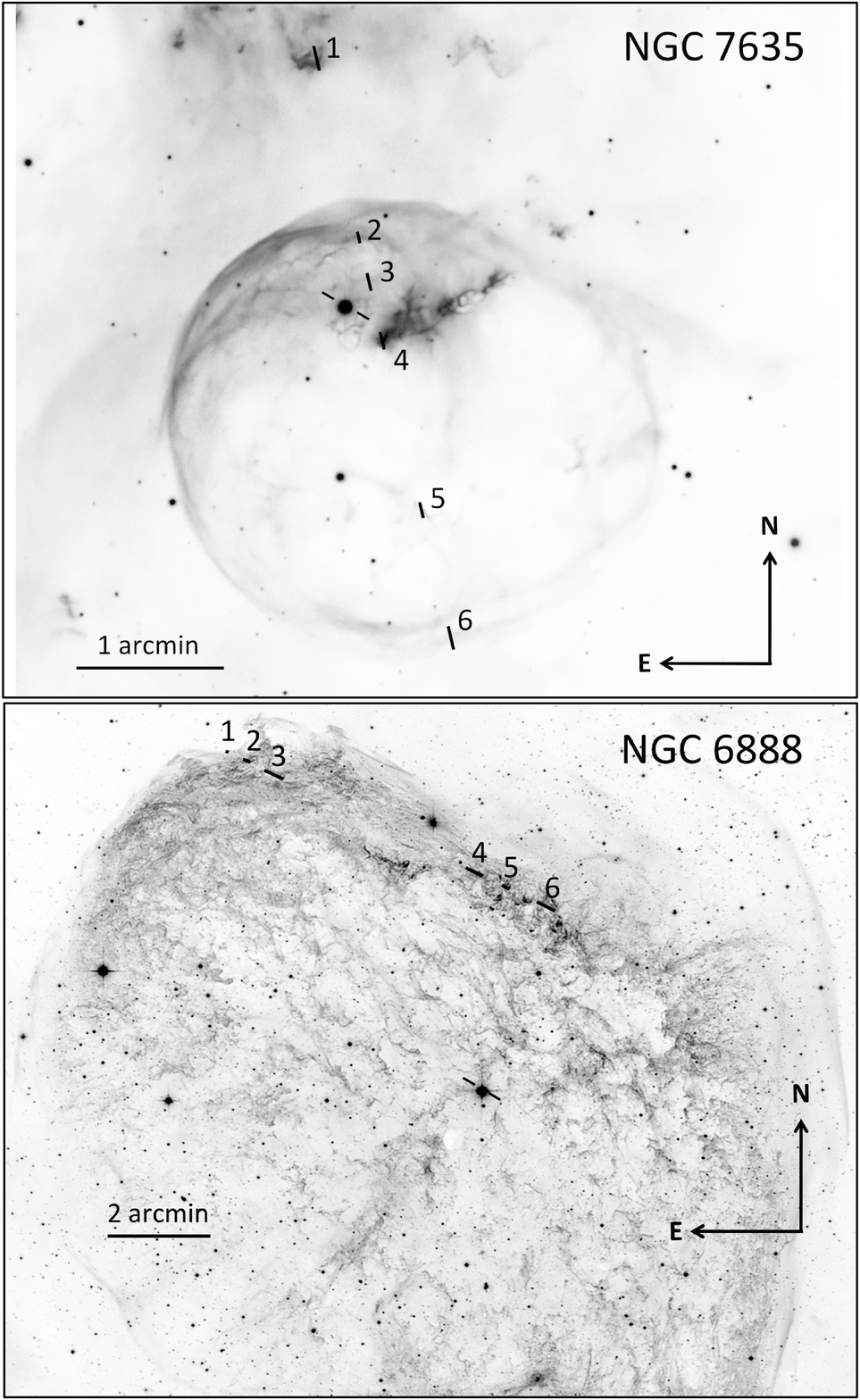} 
   \caption{Apertures observed in NGCC~7635 and NGC~6888. The image of NGC~7635 is adapted from an original one based on a combination of H$\alpha$, {\foiii} and {\fsii}   
   emission exposures taken by Josef P\"ousel and Stefan Binnewies (www.capella-observatory.com). The image of NGC~6888 is adapted from a combination of H$\alpha$ and {\foiii} emission exposures taken by 
   Daniel L\' opez (IAC/INT). The position of the ionising stars of the nebulae are indicated by a pair of short lines.}
   \label{chart_2}
  \end{figure}
 %%%%%%%%%%%%%%% 

%%%%%%%%%%%%%
  \begin{table*}
   \centering
   \begin{minipage}{150mm}
     \caption{Journal of observations.}
     \label{tab:log}
    \begin{tabular}{lccccccccc}
     \hline
     	&  & & & & & & & Exposure \\ 
     	& Ionising & Spectral & Telescope/& & & & PA & Time \\
       Nebula &  Star &Type$^{\rm a}$ & /Spectrograph & RA$^{\rm b}$ & DEC$^{\rm b}$ & ($^\circ$) &  Grating$^{\rm c}$ &  (s)\\
     \hline
       S~308 & WR6 & WN4 & Magellan/MagE & 06 53 02.2 & $-$23 53 30 & 108$^{\rm d}$ & $-$ & 1800 \\ 
       RCW~52 & CPD$-$57 3781 & O8 &Magellan/MagE & 10 46 02.5 & $-$58 38 05 & 28$^{\rm d}$ & $-$ & 4400 \\
       RCW~58 & WR40 & WN8 & Magellan/MagE & 11 06 00.1 & $-$65 32 06 & $-$32$^{\rm d}$ & $-$ & 6000 \\
       G2.4+1.4 & WR102 & WO2 & Magellan/MagE & 17 45 53.9 & $-$26 09 57 & $-$103$^{\rm d}$ & $-$ & 6000 \\
       & & &GTC/OSIRIS & 17 45 54.4 & $-$26 08 49 & 4& R1000B & 60, 4460  \\
       & & & & & & & R2500V & 60, 3568 \\
       NGC~6888 & WR136 & WN6 & GTC/OSIRIS & 20 12 29.6 & +38 27 23 & 66 & R1000B & 60, 4455 \\
       & & & & & & & R2500V & 60, 3564 \\
       NGC~7635 & BD$+$60 2522 & O6.5f & GTC/OSIRIS & 23 20 48.5 & +61 12 00 & 13 & R1000B & 60, 4455 \\
       & & & & & & & R2500V & 60, 3564 \\
            \hline
    \end{tabular}
    \begin{description}
      \item[$^{\rm a}$] Of the ionising star. 
      \item[$^{\rm b}$] Coordinates of the slit center (J2000.0).  
      \item[$^{\rm c}$] MagE has a single available grating.
      \item[$^{\rm d}$] Parallactic angle.
    \end{description}
   \end{minipage}
  \end{table*}
 %%%%%%%%%%%%%

The absence of bright emission lines of C in the optical spectrum of ionised gas has hampered to determine the C abundance in ring nebulae. This is of special importance because C is one of the key elements of the CNO cycle and its abundance should be affected by nucleosynthesis in the H-burning zone. \citet{rodriguez99, mooreetal02b} and \citet{mesadelgadoesteban10} measured the {\cii} 4267 \AA\ recombination line (hereafter RL) in NGC~7635 and \citet{mesadelgadoetal14} in NGC~6888. 
In particular, the \ionic{C}{2+}/\ionic{H}{+} ratios determined by \citet{mesadelgadoesteban10}, \citet{mesadelgadoetal14} and \citet{vamvatira-nakouetal16} led to C abundances substantially higher than those expected by the Galactic C radial abundance gradient, a result that is inconsistent with nucleosynthesis predictions. On the other hand, \citet{mesadelgadoesteban10} measured the \oii\ RLs around 4650 \AA\ in an integrated spectrum of NGC~7635 finding that the \ionic{O}{2+}/\ionic{H}{+} ratio 
determined from RLs was about 0.6 dex higher that that determined from the far much brighter {\foiii} collisionally excited lines (hereafter CELs). This is the so-called abundance discrepancy problem and affects virtually to all ionised nebulae. However, the difference found in NGC~7635 is much higher than the values usually found in {\hii} regions \citep{garciarojasesteban07}. 

The aim of this paper is to use high-efficiency spectrographs in large-aperture telescopes in order to reassess the chemical abundances of several of the brightest and more interesting ring nebulae surrounding Galactic WR and O-type massive stars. The use of this state-of-the-art instrumentation would permit to measure {\cii} and/or {\oii} RLs in some of the objects and therefore to derive the C/H ratio and explore the behavior of 
the abundance discrepancy problem in ring nebulae. Another aim of the work is to compare the abundance pattern of the stellar ejecta objects with predictions of stellar evolution models, applying the same procedure followed by \citet{mesadelgadoetal14} in order to constrain the mass of their progenitor stars. 

The structure of the paper is as follows. In Section~\ref{sec:obs}, we describe the observations and the data reduction procedure. In Section~\ref{sec:lines}, we describe the emission-line measurements and identifications as well as the reddening correction. In Section~\ref{sec:results}, we present the physical conditions, ionic and total abundances determined for the sample objects, and the abundance discrepancy factors and temperature fluctuations parameter values we obtain for some of the nebulae. In Section~\ref{sec:discussion}, we discuss the behavior of the abundance patterns found in the sample nebulae, the presence of shock excitation in localized zones of NGC~6888 and the comparison with stellar evolution models. Finally, in Section~\ref{sec:conclusions}, we summarize our main conclusions.

%%%%%%%%%%%%%%%%%%%%%%%%%%%%%%%%%%%%%%%%%%%%%%%%%%%%%%%%%%%%%%%%%%%%%%%%%%%%%%%%%%%%%%%%%%%%%%%%%
\section{Observations and Data Reduction}
\label{sec:obs}
%%%%%%%%%%%%%%%%%%%%%%%%%%%%%%%%%%%%%%%%%%%%%%%%%%%%%%%%%%%%%%%%%%%%%%%%%%%%%%%%%%%%%%%%%%%%%%%%%

The observations were performed using two telescopes, one in each hemisphere: the 10.4m Gran Telescopio Canarias (GTC) located at  
Observatorio del Roque de los Muchachos (La Palma, Spain) and the 6.5 Magellan Clay telescope at Las Campanas Observatory (Chile).

 %%%%%%%%%%%%%%% 
  \begin{figure*}
   \centering
   \includegraphics[width=177mm]{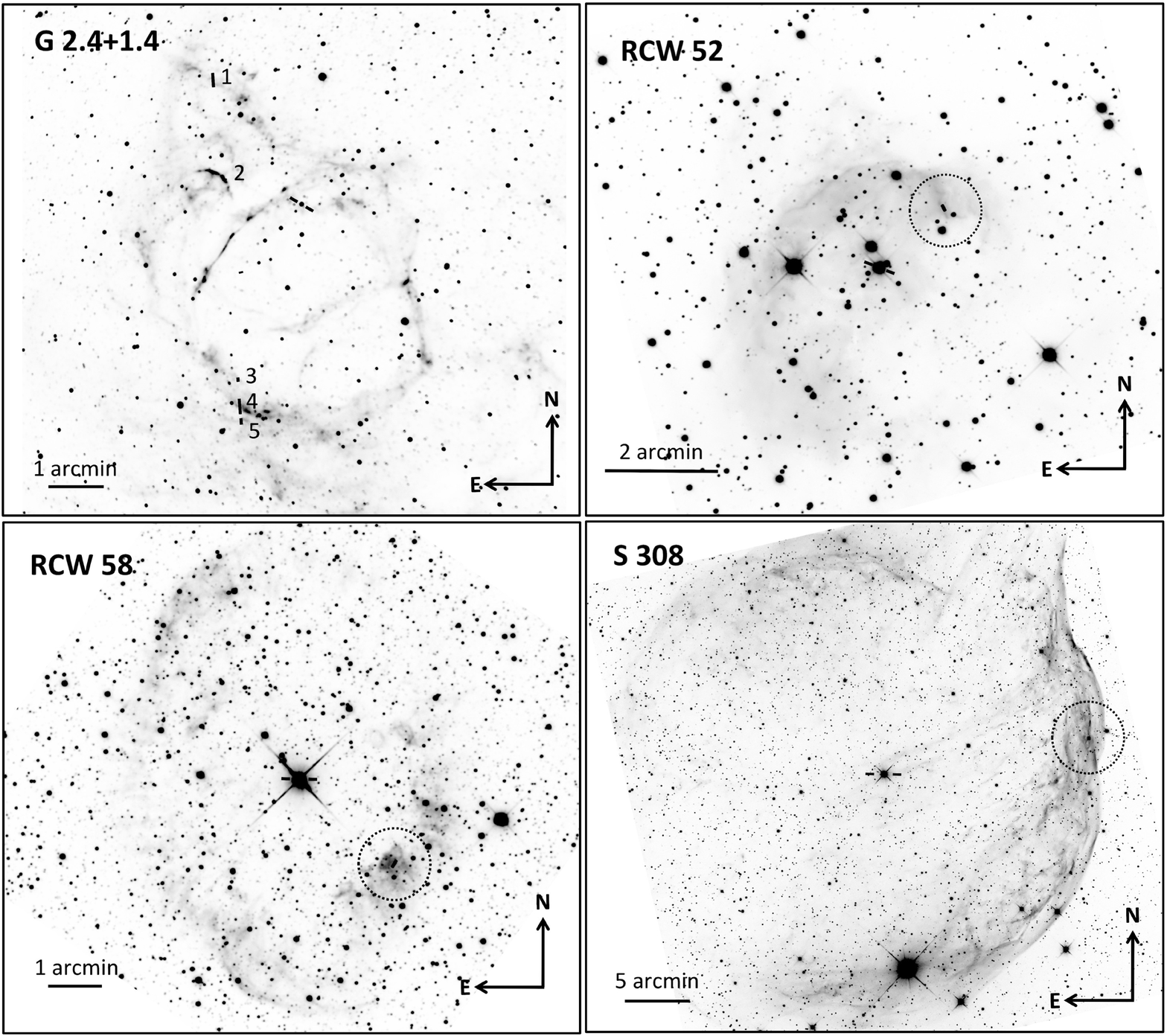} 
   \caption{Observed slit positions over G2.4+1.4, RCW~52, RCW~58 and S~308. The centre of the circles drawn over RCW~52, RCW~58 and S~308 indicate the slit positions in these objects.  
   The image of G2.4+1.4 is adapted from an original H$\alpha$ exposure taken with the IAC80 telescope at Observatorio del 
   Teide. The images of RCW~52 and RCW~58 are based on original H$\alpha$ and RGB exposures obtained by Rick Gilbert (Star Shadows Remote Observatory, SSRO). The image of S~308  
   is adapted from a combination of H$\alpha$ and {\foiii} emission exposures taken by Don Goldman (www.astrodonimaging.com). The position of the ionising stars of the nebulae are indicated by a pair of 
   short lines.}
   \label{chart_4}
  \end{figure*}
 %%%%%%%%%%%%%%% 

Spectra of NGC~7635, NGC~6888 and G2.4+1.4 were taken with OSIRIS (Optical System for Imaging and low-Intermediate-Resolution 
Integrated Spectroscopy) spectrograph \citep{cepaetal00, cepaetal03} attached to the GTC telescope. OSIRIS consists of a mosaic of 
two Marconi CCD42-82 CCDs each with 2048 $\times$ 4096 pixels and a 74 pixel gap between them. Each pixel has a physical size of 15$\mu$m. We 
used a binning 2 $\times$ 2 for our observations, giving a plate scale of 0.254 arcsec. OSIRIS was used in long-slit mode. The slit length was 7.4 arcmin and the width was set at 0.8 arcsec. 
We used two grisms in the OSIRIS observations, R1000B and R2500V. R1000B was used to cover the whole optical spectral range, from 3640 to 7870 \AA. 
R2500V covered from 4430 to 6070 \AA\ and was used to improve the resolution in the spectral zone where \foiii\ 4363 \AA\ or the RLs of \oii\ around 4650 \AA\ lie. 
The effective spectral resolution was 6.52 and 2.46 \AA\ for the R1000B and R2500V grisms, respectively. The slit was placed covering the brightest and/or interesting zones of the nebulae. 
Table~\ref{tab:log} shows the coordinates of the centre and position angle (PA) of the slit, and total integration time of the spectra. We selected several apertures in the objects observed with OSIRIS in long-slit mode (NGC~6888, NGC~7635 and G2.4+1.4) in order to extract one-dimensional spectra. These zones were selected attending to their brightness, 
particular location and the detection of the auroral collisionally excited lines \foiii\ 4363 \AA\ or \fnii\ 5755 \AA\ that are used for the determination of electron temperature. 
In figures~\ref{chart_2} and \ref{chart_4} we indicate the position of the areas -- apertures -- extracted for each object. For example, in the case of NGC~7635, our aperture no. 1 was placed in order to determine the 
physical conditions and abundances of the neighboring diffuse {\hii} region  S~162.
The size of the different apertures is indicated in the third to last row of tables~\ref{linesNGC7635}, \ref{linesNGC6888}, \ref{linesG24} and \ref{linesrest}. The OSIRIS spectra were wavelength calibrated with Hg-Ar, Ne and Xe lamps. 
The correction for atmospheric extinction was performed using the average curve for continuous atmospheric extinction at Roque de los Muchachos Observatory. The absolute 
flux calibration was achieved by observations of the standard stars BD+33~2642, GD248 \citep{oke90}, Hiltner 600 \citep{masseyetal88} and Ross~640 \citep{oke74}.  We used the {\sc iraf}\footnote{{\sc iraf} is distributed by National Optical Astronomy Observatory, which is operated by Association of Universities for Research in Astronomy, under cooperative agreement with the National Science Foundation.}{\sc twodspec} reduction package to perform bias correction, flat-fielding, cosmic-ray rejection, sky emission subtraction, wavelength and flux calibration. We checked the relative flux calibration between the bluest and the reddest wavelengths by calibrating with the different spectrophotometric standards observed in the same night and with each standard itself. The error in the relative flux calibration along the spectral range covered was lower than 3\%.

Two half-nights were granted in Las Campanas Observatory (Chile) on 30th January and 13th May 2013 to observe four ring nebulae of the southern hemisphere. On the one hand, RCW~52 and S~308 were observed during the first night that corresponded to bright time, so the spectroscopic data presented features of scattered moonlight. On the other hand, G$2.4+1.4$ and RCW~58 were observed in the second night during dark time. The observations have provided the deepest spectra available for these objects so far and were obtained using the Magellan Echellette (MagE) spectrograph in the 6.5m Clay Telescope \citep{marshalletal08}. Within the same chip, this instrument provides echelle spectroscopy on the entire optical range (3100 -- 10000 \AA) with a slit of 10 arcsec long and a fixed plate scale of 0.3 arcsec. The slit width was set to 1.0 arcsec, providing a spectral resolution of about $\lambda/\Delta\lambda\sim4000$. Each target was observed in a single slit position, placed on the sky along the parallactic angle and covering the brightest areas of the nebulae. The
MagE spectra were reduced making use of the {\sc echelle} package of {\sc iraf}, following the typical reduction process: bias subtraction, order tracing, aperture extractions, flat-fielding, wavelength calibration and flux calibration. The final spectra of RCW~58, RCW~52 and S~308 were extracted for areas of about 8-9 arcsec$^2$, using most of the slit length. In the case of G$2.4+1.4$, an area of 4 arcsec$^2$ was extracted and centered on a bright knot to maximize the signal-to-noise ratio of the final spectra. The wavelength calibration was derived from Th-Ar arcs observed after and before each target. To achieve the flux calibration of the dataset, the spectrophotometric standard star GD~108 \citep{oke90} was observed on 30th January, while GD~108 and EG~274 \citep{hamuyetal94} were observed on 13th May. Though the seeing was about 0.8 arcsec, the nights were not completely photometric. We applied the same procedure than for the GTC data to estimate the quality of the flux calibration. The error in the relative flux calibration was lower than 3\%.

%%%%%%%%%%%%%%%%%%%%%%%%%%%%%%%%%%%%%%%%%%%%%%%%%%%%%%%%%%%%%%%%%%%%%%%%%%%%%%%%%%%%%%%%%%%%%%%%%
\section{Line intensity measurements}
\label{sec:lines}
%%%%%%%%%%%%%%%%%%%%%%%%%%%%%%%%%%%%%%%%%%%%%%%%%%%%%%%%%%%%%%%%%%%%%%%%%%%%%%%%%%%%%%%%%%%%%%%%%

\begin{figure}
   \centering
   \includegraphics[scale=0.45]{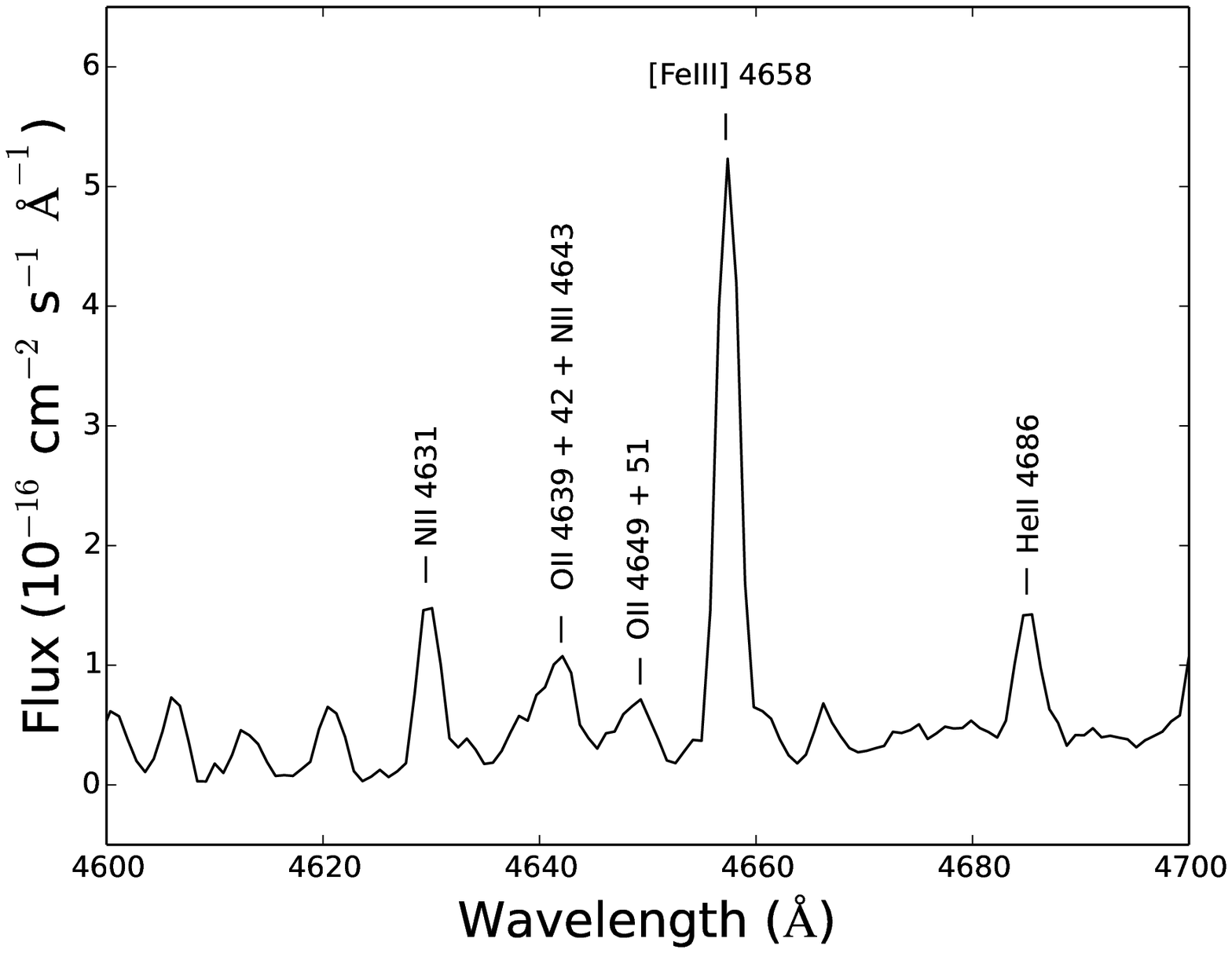} 
   \caption{Section of the OSIRIS spectrum of aperture no. 6 of NGC~6888 --uncorrected for reddening-- showing the recombination lines of multiplet 1 of  {\oii} about 4650 \AA\ and other 
   nearby emission lines.}
   \label{NGC6888_A6_OII}
  \end{figure}

Line fluxes were measured with the {\sc splot} routine of {\sc iraf} by integrating all the flux in the line between two given 
limits and over the average local continuum. All line fluxes of a given spectrum have been normalized to H$\beta$ = 100.0. In general, in the case of GTC data, the lines measured in the spectra taken
with the grating 2500V were used instead of those measured with the 1000B in the spectral range between 4430 and 6070 \AA. 
 In the case of line blending, we applied a double or multiple Gaussian profile fit procedure using {\sc splot} routine of {\sc iraf} to measure the 
individual line intensities. Line deblending was especially important in the spectral zone around multiplet 1 of \oii\ at about 4650 \AA. In the case of NGC~6888, 
where the signal-to-noise ratio of the spectra is much higher and the line profiles better resolved (see Figure~\ref{NGC6888_A6_OII}), we used Starlink {\sc dipso} software for the Gaussian fitting \citep{howardmurray90} to deblend the lines
 in the spectral range around 4650 \AA. For each 
single or multiple Gaussian fit, {\sc dipso} yields the fit parameters  (centroid, gaussian sigma, flux, FWHM, etc.) and their associated statistical errors. The identification of the lines were obtained following our 
previous works on spectroscopy of bright Galactic \hii\ regions \citep[see][and references therein]{garciarojasesteban07}. 

 %%%%%%%%%%%%%%% 
 %%%%%%%%%%%%%%% 

The reddening coefficient, $c$({\hb}), was determined from the comparison of the observed flux 
 ratio of the brightest Balmer lines -- H$\alpha$, H$\gamma$ and H$\delta$ -- with respect to H$\beta$ and the theoretical line ratios computed 
 by \cite{storeyhummer95} for the physical conditions of the nebulae and following an iterative process. We have used 
 the reddening function, $f(\lambda)$, normalized to H$\beta$ derived by \cite{cardellietal89} and assuming $R_V$ = 3.1. 
 
In tables~\ref{linesNGC7635}, \ref{linesNGC6888}, \ref{linesG24} and \ref{linesrest} we include the list of line identifications -- first 3 columns,  the reddening function, $f(\lambda)$ -- fourth column -- and dereddened flux line ratios 
with respect to H$\beta$ -- rest of columns. The quoted line intensity errors include the uncertainties in line 
flux measurement and error propagation in the reddening coefficient. The 3 last rows of each table include the size, the reddening coefficient and the observed --uncorrected for reddening-- 
integrated H$\beta$ flux, $F$(H$\beta$), of each aperture. 

%leemos la tabla de intensidad de lineas de NGC7635
  \setcounter{table}{1}
   \begin{table*}
  \centering
   \begin{minipage}{180mm}
     \caption{Derredened line intensity ratios with respect to $I$(\hb) = 100 for NGC~7635.}
     \label{linesNGC7635}
\begin{tabular}{lccccccccc}
     \hline                                              
        $\lambda_0$ & & & & \multicolumn{6}{c}{$I$($\lambda$)/$I$(\hb)} \\  
       (\AA) & Ion & ID& f($\lambda$) & A1 & A2 & A3 & A4 & A5 & A6 \\ 
     \hline        
3705	&	{\hei}	&	25	&	0.260 & $-$		&		$-$		&	4.1		:	&	0.64		:	&		$-$		&		$-$		\\
3712	&	{\hi}	&	H15	&	0.259 & $-$		&		$-$		&		$-$		&	0.87	$\pm$	0.32	&		$-$		&		$-$		\\
3726$^{\rm a}$	&	{\foii}	&	1F	&	0.257 & 192.7	$\pm$	5.8	&	91.6	$\pm$	3.8	&	128.8	$\pm$	5.0	&	192.9	$\pm$	4.1	&	162	$\pm$	12	&	151	$\pm$	11	\\
3729	&	{\foii}	&	1F	&	 &			&				&				&				&				&				\\
3750	&	{\hi}	&	H12	&	0.253 & 2.1		:	&	1.2		:	&	1.2		:	&	2.18	$\pm$	0.15	&		$-$		&		$-$		\\
3771	&	{\hi}	&	H11	&	0.249 & 2.46	$\pm$	0.78	&	3.6		:	&	3.3	$\pm$	1.1	&	3.29	$\pm$	0.20	&		$-$		&		$-$		\\
\end{tabular}
\end{minipage}
\end{table*}
%%%%%%%%%%%%%%%%%%%%%
\setcounter{table}{1}
   \begin{table*}
  \centering
   \begin{minipage}{180mm}
     \caption{continued}
    \begin{tabular}{lccccccccc}
     \hline  
        $\lambda_0$ & & & & \multicolumn{6}{c}{$I$($\lambda$)/$I$(\hb)} \\  
       (\AA) & Ion & ID& f($\lambda$) & A1 & A2 & A3 & A4 & A5 & A6 \\ 
     \hline        
4881	&	{\ffeiii}	&	2F	&	$-$0.005 & $-$		&		$-$		&	0.07		:	&	0.146	$\pm$	0.032	&		$-$		&		$-$		\\
4890	&	{\ffeii}	&	4F	&	$-$0.007 &	$-$		&		$-$		&	0.08	$\pm$	0.03	&	0.018		:	&		$-$		&		$-$		\\
4905	&	{\ffeii}	&	20F	&	$-$0.011 &	$-$		&	0.09		:	&	0.09	$\pm$	0.03	&		$-$		&		$-$		&		$-$		\\
4922	&	{\hei}	&	48	&	$-$0.015 & 0.94	$\pm$	0.06	&	1.34	$\pm$	0.11	&	1.36	$\pm$	0.11	&	0.867	$\pm$	0.022	&	1.05	$\pm$	0.22	&	0.76	$\pm$	0.17	\\
4931	&	{\ffeiii}	&	1F	&	$-$0.017 &	$-$		&		$-$		&		$-$		&	0.018	$\pm$	0.005	&		$-$		&		$-$		\\
4931	&	{\foiii}	&	1F	&		&		&				&				&				&				&				\\
\end{tabular}
\end{minipage}
\end{table*}
%%%%%%%%%%%%%%%%%%%%%
\setcounter{table}{1}
   \begin{table*}
  \centering
   \begin{minipage}{180mm}
     \caption{continued}
    \begin{tabular}{lccccccccc}
     \hline  
        $\lambda_0$ & & & & \multicolumn{6}{c}{$I$($\lambda$)/$I$(\hb)} \\  
       (\AA) & Ion & ID& f($\lambda$) & A1 & A2 & A3 & A4 & A5 & A6 \\ 
     \hline        
6548	&	{\fnii}	&	1F	&	$-$0.318 & 28.9	$\pm$	1.1	&	10.32	$\pm$	0.76	&	13.49	$\pm$	0.43	&	40.34	$\pm$	0.90	&	19.8	$\pm$	1.1	&	20.16	$\pm$	0.78	\\
6563	&	{\hi}	&	{\ha}	&	$-$0.320 & 281.5	$\pm$	7.8	&	292.0	$\pm$	6.8	&	306.9	$\pm$	9.9	&	287.7	$\pm$	6.2	&	325	$\pm$	18	&	309	$\pm$	11	\\
6583	&	{\fnii}	&	1F	&	$-$0.323 & 85.9	$\pm$	2.4	&	32.43	$\pm$	0.78	&	41.3	$\pm$	1.3	&	123.2	$\pm$	2.7	&	59.9	$\pm$	3.4	&	59.7	$\pm$	2.2	\\
6678	&	{\hei}	&	46	&	$-$0.336 & 3.13	$\pm$	0.09	&	4.72	$\pm$	0.15	&	4.04	$\pm$	0.17	&	2.860	$\pm$	0.064	&	3.64	$\pm$	0.23	&	3.26	$\pm$	0.21	\\
6716	&	{\fsii}	&	2F	&	$-$0.342 & 26.06	$\pm$	0.74	&	6.67	$\pm$	0.17	&	9.19	$\pm$	0.32	&	13.74	$\pm$	0.30	&	13.23	$\pm$	0.79	&	15.84	$\pm$	0.64	\\
6731	&	{\fsii}	&	2F	&	$-$0.344 & 27.12	$\pm$	0.77	&	5.23	$\pm$	0.14	&	6.96	$\pm$	0.25	&	18.83	$\pm$	0.41	&	10.94	$\pm$	0.67	&	11.36	$\pm$	0.47	\\
\hline                                              
    \end{tabular}
    \begin{description}
      \item[$^{\rm a}$] Blend of {\foii} 3726 and 3729 \AA\ lines.  
      \item[$^{\rm b}$] Blend of {\foii} 7318 and 7320 \AA\ lines and affected by sky emission.  
      \item[$^{\rm c}$] Blend of {\foii} 7330 and 7331 \AA\ lines and affected by sky emission.                   
    \end{description}
   \end{minipage}
   \end{table*}

  %%%%%%%%%%%%%%%%
 
  %leemos la tabla de intensidad de lineas de NGC6888
  \setcounter{table}{2}
   \begin{table*}
  \centering
   \begin{minipage}{180mm}
     \caption{Derredened line intensity ratios with respect to $I$(\hb) = 100 for NGC~6888.}
     \label{linesNGC6888}
    \begin{tabular}{lccccccccc}
     \hline                                              
        $\lambda_0$ & & & & \multicolumn{6}{c}{$I$($\lambda$)/$I$(\hb)} \\  
       (\AA) & Ion & ID& f($\lambda$) & A1 & A2 & A3 & A4 & A5 & A6 \\ 
     \hline        
3705	&	{\hei}	&	25	&	0.260	&				&		$-$		&		$-$		&	2.5		:	&		$-$		&		$-$		\\
3726$^{\rm a}$	&	{\foii}	&	1F	&	0.257	&	256.4	$\pm$	5.3	&	124.1	$\pm$	5.9	&	93.9	$\pm$	5.0	&	45.4	$\pm$	4.6	&	42.6	$\pm$	2.2	&	50.4	$\pm$	4.7	\\
3729	&	{\foii}	&	1F	&		&				&				&				&				&				&				\\
3750	&	{\hi}	&	H12	&	0.253	&		$-$		&	1.7		:	&	1.7		:	&	1.6		:	&	1.66	$\pm$	0.51	&	2.1		:	\\
3771	&	{\hi}	&	H11	&	0.249	&		$-$		&	3.6	$\pm$	1.1	&	3.1		:	&	3.3	$\pm$	1.2	&	3.33	$\pm$	0.70	&	3.6		:	\\
3798	&	{\hi}	&	H10	&	0.244	&		$-$		&	3.77	$\pm$	0.84	&	3.7		:	&	4.0	$\pm$	1.4	&	3.71	$\pm$	0.77	&	4.3	$\pm$	1.6	\\
\end{tabular}
\end{minipage}
\end{table*}
%%%%%%%%%%%%%%%%%%%%%
\setcounter{table}{2}
   \begin{table*}
  \centering
   \begin{minipage}{180mm}
     \caption{continued}
    \begin{tabular}{lccccccccc}
     \hline  
        $\lambda_0$ & & & & \multicolumn{6}{c}{$I$($\lambda$)/$I$(\hb)} \\  
       (\AA) & Ion & ID& f($\lambda$) & A1 & A2 & A3 & A4 & A5 & A6 \\ 
     \hline   
5159	&	{\ffeii}	&	19F	&	-0.073	&		$-$		&		$-$		&	0.118	$\pm$	0.034	&		$-$		&		$-$		&		$-$		\\
5192	&	{\fariii}	&	3F	&	-0.081	&		$-$		&	0.06		:	&	0.074		:	&	0.05		:	&	0.045	$\pm$	0.013	&	0.058		:	\\
5270	&	{\ffeiii}	&	1F	&	-0.100	&		$-$		&	0.498	$\pm$	0.056	&	0.920	$\pm$	0.073	&	0.755	$\pm$	0.053	&	0.842	$\pm$	0.063	&	0.644	$\pm$	0.054	\\
5412	&	{\ffeiii}	&	1F	&	-0.134	&		$-$		&	0.08		:	&	0.130	$\pm$	0.026	&	0.111	$\pm$	0.033	&	0.089	$\pm$	0.028	&	0.074		:	\\
5412	&	{\heii}	&	4.7	&		&				&				&				&				&				&				\\
5518	&	{\fcliii}	&	1F	&	-0.154	&	1.30	$\pm$	0.45	&	0.495	$\pm$	0.055	&	0.472	$\pm$	0.050	&	0.492	$\pm$	0.043	&	0.542	$\pm$	0.032	&	0.529	$\pm$	0.053	\\
\hline                                              
    \end{tabular}
    \begin{description}
      \item[$^{\rm a}$] Blend of {\foii} 3726 and 3729 \AA\ lines.  
      \item[$^{\rm b}$] Blend of {\foii} 7318 and 7320 \AA\ lines and affected by sky emission.  
      \item[$^{\rm c}$] Blend of {\foii} 7330 and 7331 \AA\ lines and affected by sky emission.                      
    \end{description}
   \end{minipage}
   \end{table*}

  %%%%%%%%%%%%%%%%

  %leemos la tabla de intensidad de lineas de G2.4+1.4
     \begin{table*}
  \centering
   \begin{minipage}{180mm}
     \caption{Derredened line intensity ratios with respect to $I$(\hb) = 100 for G2.4+1.4.}
     \label{linesG24}
    \begin{tabular}{lcccccccc}
     \hline                                              
        $\lambda_0$ & & & & \multicolumn{5}{c}{$I$($\lambda$)/$I$(\hb)} \\  
       (\AA) & Ion & ID& f($\lambda$) & A1 & A2 & A3 & A4 & A5  \\ 
     \hline     
3726.03$^{\rm a}$	&	{\foii}	&	1F	&	0.257	&	184	$\pm$	46	&	115.7	$\pm$	6.2	&	219		:	&	96	$\pm$	27	&	208	$\pm$	62	\\
3728.82	&	{\foii}	&	1F	&	0.256	&				&	92.7	$\pm$	5.8	&				&				&				\\
3835.39	&	{\hi}	&	H9	&	0.237	&		$-$		&	8.6	$\pm$	2.8	&		$-$		&		$-$		&		$-$		\\
3867.49	&	{\hei}	&	20	&	0.230	&	39		:	&	67.9	$\pm$	5.9	&		$-$		&	57	$\pm$	13	&	54		:	\\
3868.75	&	{\fneiii} 	&	1F	&		&				&				&				&				&				\\
3888.65	&	{\hei}	&	5	&	0.227	&		$-$		&	16.7	$\pm$	3.4	&		$-$		&	12 		:	&		$-$		\\
\hline                                              
    \end{tabular}
    \begin{description}
      \item[$^{\rm a}$] Blend of {\foii} 3726 and 3729 \AA\ lines.  
      \item[$^{\rm b}$] Blend of {\foii} 7318 and 7320 \AA\ lines and affected by sky emission.  
      \item[$^{\rm c}$] Blend of {\foii} 7330 and 7331 \AA\ lines and affected by sky emission.    
    \end{description}
   \end{minipage}
   \end{table*}

  %%%%%%%%%%%%%%%%

%leemos la tabla de intensidad de lineas de RCW~52, RCW~58 y S308
    \begin{table*}
  \centering
   \begin{minipage}{180mm}
     \caption{Derredened line intensity ratios with respect to $I$(\hb) = 100 for RCW~52, RCW~58 and S~308.}
     \label{linesrest}
    \begin{tabular}{lcccccc}
     \hline                                              
        $\lambda_0$ & & & & \multicolumn{3}{c}{$I$($\lambda$)/$I$(\hb)} \\  
       (\AA) & Ion & ID& f($\lambda$) & RCW~52 & RCW~58 & S~308 \\ 
     \hline  
 3726.03	&	{\foii}	&	1F	&	0.257	&	168.5	$\pm$	5.6	&	24.2	$\pm$	1.2	&	33		:	\\
3728.82	&	{\foii}	&	1F	&	0.256	&	209.0	$\pm$	7.0	&	33.5	$\pm$	1.4	&	52	$\pm$	16	\\
3734.37	&	{\hi}	&	H13	&	0.255	&		$-$		&	2.19	$\pm$	0.46	&		$-$		\\
3750.15	&	{\hi}	&	H12	&	0.253	&		$-$		&	4.14	$\pm$	0.98	&		$-$		\\
3770.63	&	{\hi}	&	H11	&	0.249	&		$-$		&	3.88	$\pm$	0.62	&		$-$		\\
3797.9	&	{\hi}	&	H10	&	0.244	&		$-$		&	3.59	$\pm$	0.35	&		$-$		\\
\hline                                              
    \end{tabular}
    \begin{description}
      \item[$^{\rm a}$] Blend of {\foii} 7318 and 7320 \AA\ lines and affected by sky emission.  
      \item[$^{\rm b}$] Blend of {\foii} 7330 and 7331 \AA\ lines and affected by sky emission.                   
    \end{description}
   \end{minipage}
   \end{table*}

  %%%%%%%%%%%%%%%%

 %%%%%%%%%%%%%
  \begin{table*}
   \centering
   \begin{minipage}{180mm}
   \caption{Physical Conditions, ionic and total abundances for the apertures extracted in NGC~7635.}
  \label{cond_abun_NGC7635}
    \begin{tabular}{l c c c c c c c c c}
     \hline
        & A1 & \multicolumn{2}{c}{A2} & \multicolumn{2}{c}{A3} & \multicolumn{2}{c}{A4} & A5 & A6 \\
     	& {\tf} = 0.0 & {\tf} = 0.0 & {\tf} = 0.061 & {\tf} = 0.0 & {\tf} = 0.049 & {\tf} = 0.0 & {\tf} = 0.041 & {\tf} = 0.0 & {\tf} = 0.0  \\
	& &  &  $\pm$   0.016 & &  $\pm$   0.017 & &  $\pm$   0.021 & &  \\
    \hline
	\multicolumn{10}{c}{Physical conditions$^{\rm a}$} \\
	\\
	{\elecd}({\fsii}) & 680  $\pm$   110 & \multicolumn{2}{c}{150  $\pm$  50} & \multicolumn{2}{c}{110  $\pm$  60} & \multicolumn{2}{c}{2140  $\pm$  290} & 220  $\pm$  130 & $<$100 \\
	{\elecd}({\fcliii}) & 240$^{+370}_{-240}$ & \multicolumn{2}{c}{220$^{+750}_{-220}$} & \multicolumn{2}{c}{$-$} & \multicolumn{2}{c}{3290  $\pm$  380} & $-$ & $-$ \\
	\\
\hline
    \end{tabular}
    \begin{description}
      \item[$^{\rm a}$] {\elecd} in {\cmc}; {\elect} in K.     
      \item[$^{\rm b}$] In units of 12+log(\ionic{X}{n+}/\ionic{H}{+}).  
      \item[$^{\rm c}$] In units of 12+log(X/H).     
    \end{description}
   \end{minipage}
  \end{table*}
 %%%%%%%%%%%%%

 %%%%%%%%%%%%%
  \begin{table*}
   \centering
   \begin{minipage}{180mm}
   \caption{Physical Conditions, ionic and total abundances for the apertures extracted in NGC~6888.}
  \label{cond_abun_NGC6888}
    \begin{tabular}{l c c c c c c c c c}
     \hline
        & A1 & A2 & A3 & \multicolumn{2}{c}{A4} & \multicolumn{2}{c}{A5} & \multicolumn{2}{c}{A6}  \\
     	& {\tf} = 0.0 & {\tf} = 0.0 &  {\tf} = 0.0 &  {\tf} = 0.0 & {\tf} = 0.83 & {\tf} = 0.0 & {\tf} = 0.071 & {\tf} = 0.0 & {\tf} = 0.069  \\
	& &  & & &  $\pm$   0.019 & &  $\pm$   0.014 & &  $\pm$   0.016  \\
    \hline
	\multicolumn{10}{c}{Physical conditions$^{\rm a}$} \\
	\\
	{\elecd}({\fsii}) & $<$100 & 310  $\pm$  90 & 250  $\pm$  150 & \multicolumn{2}{c}{150  $\pm$  90} & \multicolumn{2}{c}{210  $\pm$  140} & \multicolumn{2}{c}{200  $\pm$  140}  \\
	{\elecd}({\fcliii}) & $-$ & 85$^{+980}_{-85}$ & $-$ & \multicolumn{2}{c}{120$^{+640}_{-120}$} & \multicolumn{2}{c}{390$^{+480}_{-390}$} & \multicolumn{2}{c}{350$^{+830}_{-350}$} \\
	\\
\hline
    \end{tabular}
    \begin{description}
      \item[$^{\rm a}$] {\elecd} in {\cmc}; {\elect} in K.     
      \item[$^{\rm b}$] {\elect}(low) = {\elect}({\fnii}) in all apertures.
      \item[$^{\rm c}$] In units of 12+log(\ionic{X}{n+}/\ionic{H}{+}).  
      \item[$^{\rm d}$] In units of 12+log(X/H).        
    \end{description}
   \end{minipage}
  \end{table*}
 %%%%%%%%%%%%%

 %%%%%%%%%%%%%
  \begin{table*}
   \centering
   \begin{minipage}{180mm}
   \caption{Physical Conditions, ionic and total abundances for the apertures extracted in G2.4+1.4.}
  \label{cond_abun_G24}
    \begin{tabular}{l c c c c c}
     \hline
        & A1 & A2 & A3 & A4 & A5 \\
    \hline
	\multicolumn{6}{c}{Physical conditions$^{\rm a}$} \\
	\\
	{\elecd}({\fsii}) & 130  $\pm$  50 & 1310  $\pm$  220 &  220  $\pm$  80 & 120  $\pm$  70 &  120  $\pm$  60 \\
	{\elecd}({\foii}) & $-$ & 1130  $\pm$  270 & $-$ & $-$ & $-$ \\
	{\elecd}({\fcliii}) & $-$ & 330: & $-$ & 1810: & $-$ \\
	\\
	{\elect}({\fnii})$^{\rm b}$ & 8520  $\pm$   600 & 9060  $\pm$   340 & 11260  $\pm$  1620 & 9940  $\pm$  1220 &  8600  $\pm$  750 \\
\hline
    \end{tabular}
    \begin{description}
      \item[$^{\rm a}$] {\elecd} in {\cmc}; {\elect} in K.     
      \item[$^{\rm b}$] {\elect}(low) = {\elect}({\fnii}) in all apertures.
      \item[$^{\rm c}$] In units of 12+log(\ionic{X}{n+}/\ionic{H}{+}).  
      \item[$^{\rm d}$] In units of 12+log(X/H).        
    \end{description}
   \end{minipage}
  \end{table*}
 %%%%%%%%%%%%%

%%%%%%%%%%%%%
  \begin{table}
   \centering
   \begin{minipage}{180mm}
   \caption[caption]{Physical Conditions, ionic and total abundances\\\hspace{\textwidth} for RCW~52, RCW~58 and S~308}
  \label{cond_abun_resto}
    \begin{tabular}{l c c c}
     \hline
        & RCW~52 & RCW~58 & S~308 \\
    \hline
	\multicolumn{4}{c}{Physical conditions$^{\rm a}$} \\
	\\
	{\elecd}({\fsii}) & 280  $\pm$  150 & 100  $\pm$  100 &  $<$100 \\
	{\elecd}({\foii}) & 190  $\pm$  90 & $<$100 & $<$100 \\
	\\
	{\elect}({\fnii})$^{\rm b}$ & 7200  $\pm$   420 & 6450  $\pm$   220 & 14930  $\pm$  2900 \\
	\\
	{\elect}({\foiii}) & $-$ & $-$ & 16800  $\pm$  2500\\
	{\elect}({\fsiii}) & $-$ & $-$ & 12420  $\pm$  2200\\
	{\elect}(high) & 7200  $\pm$   420 & 6450  $\pm$   220 & 15020  $\pm$  1720\\
	\\
	\multicolumn{4}{c}{Ionic abundances$^{\rm c}$ from collisionally excited lines} \\
	\\
	N$^+$ & 7.83  $\pm$  0.07 & 8.41  $\pm$  0.06 & 7.03  $\pm$  0.16 \\ 
	O$^+$ & 8.87  $\pm$  0.13 & 8.33  $\pm$  0.09 & 6.85  $\pm$  0.61 \\ 
	O$^{2+}$ & 7.52  $\pm$  0.19 & 7.48  $\pm$  0.08 & 7.95  $\pm$  0.10 \\ 
	Ne$^{2+}$ & $-$ & $-$ & 7.55  $\pm$  0.25  \\
	S$^+$ & 6.64  $\pm$  0.09 & 6.26  $\pm$  0.07 & 5.30  $\pm$  0.17 \\ 
	S$^{2+}$ & 6.79  $\pm$  0.15 & 6.93  $\pm$  0.09 & 6.52  $\pm$  0.10 \\ 
	Ar$^{2+}$ & 6.13  $\pm$  0.20 & 6.18  $\pm$  0.11 & 6.05  $\pm$  0.16 \\ 
	Fe$^{2+}$ & $-$ & 6.44  $\pm$  0.21 &  6.28  $\pm$  0.32  \\
	\\
	\multicolumn{4}{c}{Ionic abundances$^{\rm c}$ from recombination lines} \\
	\\
	He$^+$ & 10.88  $\pm$  0.03 & 11.24  $\pm$  0.03 & 11.20  $\pm$  0.04 \\
	\\
	\multicolumn{4}{c}{Total abundances$^{\rm d}$ and abundance ratios}\\
     	\\	
 	He &  11.13  $\pm$  0.05  &  11.46  $\pm$  0.21  &  11.21  $\pm$  0.08  \\
	O & 8.91  $\pm$  0.12 &  8.39  $\pm$  0.08 & 7.98  $\pm$  0.11 \\ 
	N & 7.97  $\pm$  0.08 &  8.60  $\pm$  0.14 &  7.83  $\pm$  0.60 \\
	Ne & $-$ &  $-$ & 7.62: \\
	S &  7.02  $\pm$  0.18 &  7.02  $\pm$  0.12 & 6.64  $\pm$  0.25 \\
	Ar &  6.42  $\pm$  0.21 &  6.44  $\pm$  0.24 & 6.15  $\pm$  0.19 \\
	Fe &  $-$ &  6.87  $\pm$  0.38 & 6.82  $\pm$  1.03 \\
     \hline
    \end{tabular}
    \begin{description}
      \item[$^{\rm a}$] {\elecd} in {\cmc}; {\elect} in K.     
      \item[$^{\rm b}$] {\elect}(low) = {\elect}({\fnii}) in all objects.
      \item[$^{\rm c}$] In units of 12+log(\ionic{X}{n+}/\ionic{H}{+}).  
      \item[$^{\rm d}$] In units of 12+log(X/H).        
    \end{description}
   \end{minipage}
  \end{table}
 %%%%%%%%%%%%%

%%%%%%%%%%%%%%%%%%%%%%%%%%%%%%%%%%%%%%%%%%%%%%%%%%%%%%%%%%%%%%%%%%%%%%%%%%%%%%%%%%%%%%%%%%%%%%%%%
\section{Results}
\label{sec:results}
%%%%%%%%%%%%%%%%%%%%%%%%%%%%%%%%%%%%%%%%%%%%%%%%%%%%%%%%%%%%%%%%%%%%%%%%%%%%%%%%%%%%%%%%%%%%%%%%%

%%%%%%%%%%%%%%%%%%%%%%%%%%%%%%%%%%%%%%%%%%%%%%%%%%%%%%%%%%%%%%%%%%%%%%%%%%%%%%%%%%%%%%%%%%%%%%%%%
\subsection{Physical Conditions } \label{conditions}
%%%%%%%%%%%%%%%%%%%%%%%%%%%%%%%%%%%%%%%%%%%%%%%%%%%%%%%%%%%%%%%%%%%%%%%%%%%%%%%%%%%%%%%%%%%%%%%%%

Physical conditions were calculated making use of the version 1.0.2 of {\sc pyneb} \citep{Luridianaetal15} in combination with the atomic data listed in Table 5 of \citet{estebanetal14}. {\sc pyneb} is an updated and more versatile version 
of the {\sc nebular} package of {\sc iraf} written in {\sc python} language.  The density-sensitive emission line ratios we have used are {\fsii}~6717/6731 for all the objects; {\fcliii}~5518/5538 for several apertures of NGC~7635, NGC~6888 and G2.4+1.4 and {\foii}~3726/3729 in the case of the higher resolution spectra taken with MagE spectrograph. The electron density is below 300 {\cmc} in all the objects and apertures except in aperture no. 2 of G2.4+1.4 
and aperture no. 4 of NGC~7635 where the values are 1200 and 2500 {\cmc}, respectively. We have assumed {\elecd}({\fsii}) as representative for all the apertures and objects. 
The temperature-sensitive emission line ratios we have used are {\fnii}~5755/(6548+6584) for all objects; 
{\fsii}~(4069+4076)/(6717+6731) for a single aperture of NGC~7635 and G2.4+1.4; {\foiii}~4363/(4959+5007) 
for NGC~6888, S~308 and several apertures of NGC~7635 and G2.4+1.4; {\fsiii}~6312/(9069+9532) for S~308 and finally {\fariii}~5192/(7136+7751) for several apertures of NGC~7635, NGC~6888 and G2.4+1.4. The  temperature-sensitive emission line ratio {\foii}~(7319+7330)/(3726+3729) was not used due to the strong contamination by sky emission features.  

Temperatures and densities were determined using an iterative procedure. Firstly, we assumed an initial {\elect} of 10,000 K to determine {\elecd} and this last value was used to calculate {\elect}.  
We assumed a two-zone approximation for the nebula estimating the representative values of the electron temperature for the zones where low and high-ionisation potential ions 
are present, {\elect}(low) and {\elect}(high). In all the cases, {\elect}({\fnii}) was taken as representative of {\elect}(low). For {\elect}(high), the weighted average of {\elect}({\foiii}) and 
{\elect}({\fariii}) -- or {\elect}({\fsiii}) when available -- was considered. For RCW~52 and RCW~58 and some apertures of NGC~7635 and G2.4+1.4, the {\foiii} 4363 \AA\ line was not detected and 
we assumed {\elect}(high) = {\elect}(low) = {\elect}({\fnii}).  The physical conditions of the different apertures and nebulae are presented in tables~\ref{cond_abun_NGC7635}, ~\ref{cond_abun_NGC6888}, 
~\ref{cond_abun_G24} and ~\ref{cond_abun_resto}.  

In general, the values of {\elecd} and {\elect} we obtain are consistent with previous determinations. In the case of NGC 7635, most of the areas of the bubble show densities between 100-200 {\cmc} 
but the aperture corresponding to S~162 and the bright knot at the southwest of the ionising star -- aperture no. 4 -- show larger values of {\elecd}. \citet{rodriguez99}, \citet{mooreetal02b} and {\citet{mesadelgadoesteban10} obtained densities 
of several thousands {\cmc} and values of {\elect}({\fnii}) between 8000 and 9000 K for the knots at the southwest of the ionising star, consistent with the {\elecd} $\approx$ 2500 {\cmc} and  {\elect}({\fnii}) = 9080 we obtain for 
our aperture no. 4. For the north edge of the bubble, \citet{mooreetal02b} and {\citet{mesadelgadoesteban10} obtain values of the physical conditions very similar to the ones we derive for apertures no. 2 and 3.

\citet{fernandezmartinetal12} and \citet{reyesperezetal15} observed zones at the north of NGC~6888 not far from our apertures no. 2 and 3 -- their Zone X1 and Position A, respectively. Those authors find very 
similar {\elecd} and {\elect}({\fnii}) values but the {\elect}({\foiii}) obtained by \citet{fernandezmartinetal12} is much lower than ours. They found {\elect}({\foiii}) = 6350 K, and we obtain temperatures between 9960 and 12,650 K. This large difference can be explained attending to the fact that \citet{fernandezmartinetal12} cannot compute directly {\elect}({\foiii}). 
Their  {\foiii} 4363 \AA\ line it is blended with the sky feature of Hg\thinspace{\sc i} 4358 \AA, 
so their {\elect}({\foiii}) determinations are based on the application of empirical calibrations, that may be appropriate when using integrated spectra of a nebula but of doubtful 
application for individual apertures covering a small fraction of it. On the other hand, \citet{mesadelgadoetal14} and \citet{reyesperezetal15} observed zones rather close to our aperture no. 6 of NGC~6888 -- position B of \citet{reyesperezetal15}. The densities are consistent with 
our values in all cases as well as the {\elect}({\foiii}) determined by  \citet{mesadelgadoetal14}, however, the {\elect}({\fnii}) determined by \citet{reyesperezetal15} is about 3000 K higher. 
One of the most interesting result of our study is the finding of a very high value of  {\elect}({\foiii}) -- 19,000 K -- in aperture no. 1 of NGC~6888. The auroral {\foiii} 4343 \AA\ line has a rather high signal-to-noise ratio in this zone (see Figure~\ref{NGC6888_A1_OIII}), so the determination is highly confident. In her PhD Thesis, P. Mitra reported the detection of very high values of {\elect}({\foiii}) in 
rather low signal-to-noise spectra of peripheral zones of NGC~6888 \citep{mitra91}. In her slit position NO3-W, at the rim of the bubble -- rather close to our aperture no. 1 -- Mitra obtains {\elect}({\foiii}) $\approx$ 50,000 K. \citet{dufour89} interprets that such high values of the temperature cannot be produced by photoionisation but may be explained by shock models with velocities of about 100 km $^{-1}$  and incomplete recombination regions of the kind proposed by \citet{raymondetal88} for the Cygnus Loop. We will further discuss the characteristics of this particular spectrum in Section~\ref{aperture1}.

\citet{estebanetal92} were the first who determined {\elect} for G2.4+1.4, but based on a rather low signal-to-noise spectrum of a zone close to our aperture no. 2. They obtain 
{\elect}({\foiii}) $\approx$ 12,000 K, 1000-2000 K higher than our determinations. We derive {\elect}({\fnii}) for the first time in this object and in all apertures. The values of  electron density are 
in the range 100-200 {\cmc} except in the case of aperture no. 2, for which {\elecd}({\fsii}) and {\elecd({\fcliii}) give consistent high values of about 1200 {\cmc} . This high-density zone corresponds to the brightest area of the nebula. 

We obtain a fairly high value of {\elect} in S~308 -- about 15,000 K --, and determine {\elect}({\fnii}) for the first time in this object. \citet{estebanetal92} obtained {\elect}({\foiii}) between 13,700 and 18,600 K in different zones of S~308. All available determinations of {\elecd} for this object are always below the low-density limit, {\elecd} $<$ 100 {\cmc}.  Spectroscopical results of RCW~58 have been reported by \citet{kwitter84, rosamathis90} and \citet{stocketal11} but none of these works could derive {\elect} from line intensity ratios. Therefore our value of {\elect}({\fnii}) is the first direct determination of {\elect} in RCW~58. Finally, 
this work presents the first determination of physical conditions and chemical abundances in RCW~52. 

 %%%%%%%%%%%%%%% 
  \begin{figure}
   \centering
   \includegraphics[scale=0.45]{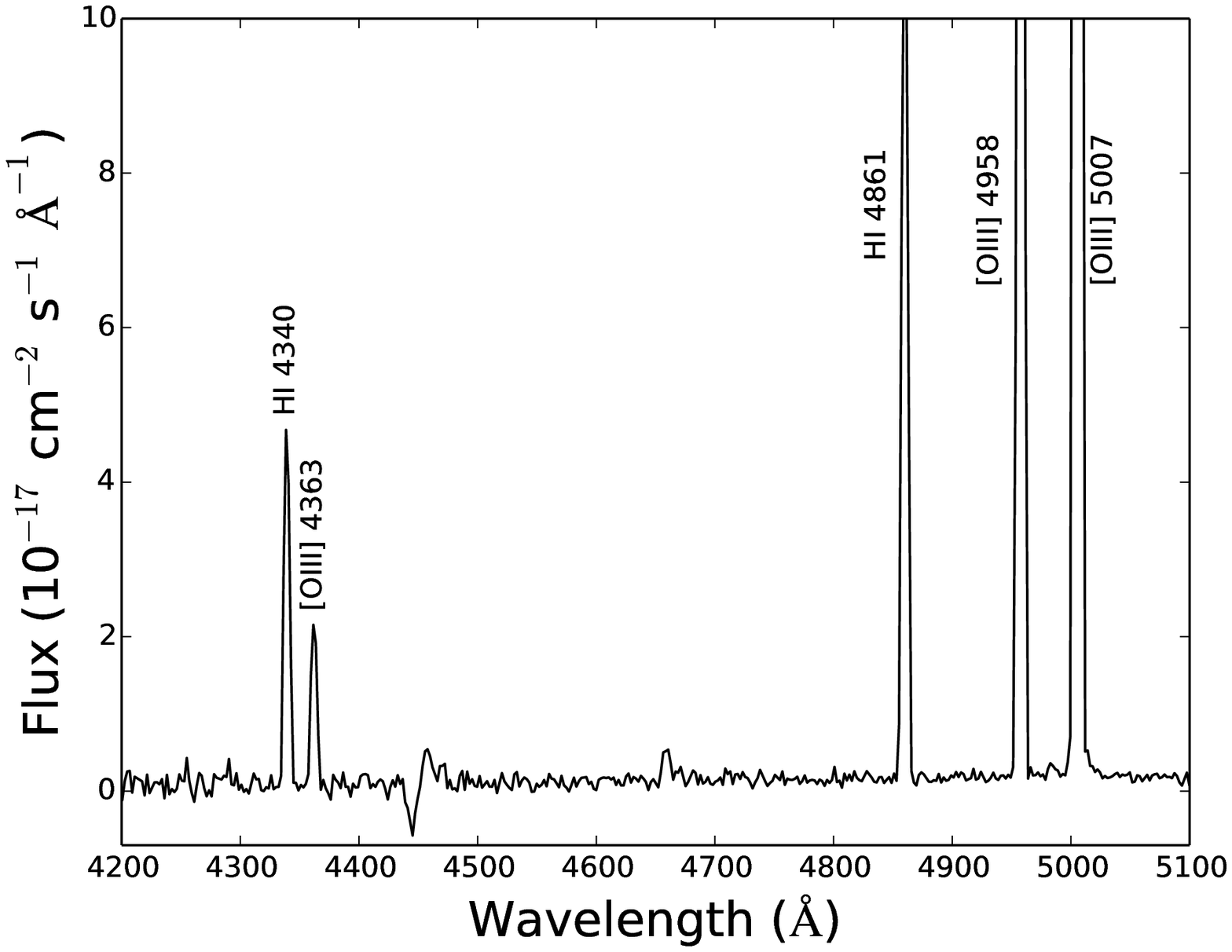} 
   \caption{Section of the OSIRIS spectrum of aperture no. 1 of NGC~6888 --uncorrected for reddening-- showing the abnormaly intense auroral {\foiii} 4363 \AA\ line in this external part of the nebula.}
   \label{NGC6888_A1_OIII}
  \end{figure}
 %%%%%%%%%%%%%%% 
 
%%%%%%%%%%%%%%%%%%%%%%%%%%%%%%%%%%%%%%%%%%%%%%%%%%%%%%%%%%%%%%%%%%%%%%%%%%%%%%%%%%%%%%%%%%%%%%%%%
\subsection{Ionic abundances from CELs } \label{cels}
%%%%%%%%%%%%%%%%%%%%%%%%%%%%%%%%%%%%%%%%%%%%%%%%%%%%%%%%%%%%%%%%%%%%%%%%%%%%%%%%%%%%%%%%%%%%%%%%%

The use of large-aperture telescopes has permitted to obtain very deep spectra of the nebulae and to determine several ionic abundances from CELs and, in some cases, also from RLs. Using CELs we have derived abundances of N$^+$, O$^+$, O$^{2+}$, S$^+$, S$^{2+}$ and Ar$^{2+}$ for all objects; Fe$^{2+}$ for all nebulae except RCW~52; Ne$^{2+}$ also for all except the low ionisation nebulae RCW~52 and RCW~58; Cl$^{2+}$ for NGC~7635, NGC~6888 and G2.4+1.4 and Ar$^{3+}$ only for G2.4+1.4. All the computations were made with {\sc pyneb} and using the atomic data listed in Table 5 of \citet{estebanetal14}. We assumed a two zone scheme adopting  {\elect}(low) for ions with low ionisation potential (N$^+$, O$^+$, S$^+$, and Fe$^{2+}$) and {\elect}(high) for the high ionisation potential ions (O$^{2+}$, Ne$^{2+}$, S$^{2+}$, Cl$^{2+}$, Ar$^{2+}$ and Ar$^{3+}$). Tables~\ref{cond_abun_NGC7635}, ~\ref{cond_abun_NGC6888}, ~\ref{cond_abun_G24} and ~\ref{cond_abun_resto} show the ionic abundances and their uncertainties, which were computed taking into account the uncertainties in line fluxes, {\elect} and {\elecd}. We have considered two sets of abundances, one for {\tf} = 0 and one for {\tf} $>$ 0, for those nebulae where the temperature fluctuations parameter, {\tf}, can be estimated (see Section~\ref{t2}).

%%%%%%%%%%%%%%% 
  \begin{figure}
   \centering
   \includegraphics[scale=0.45]{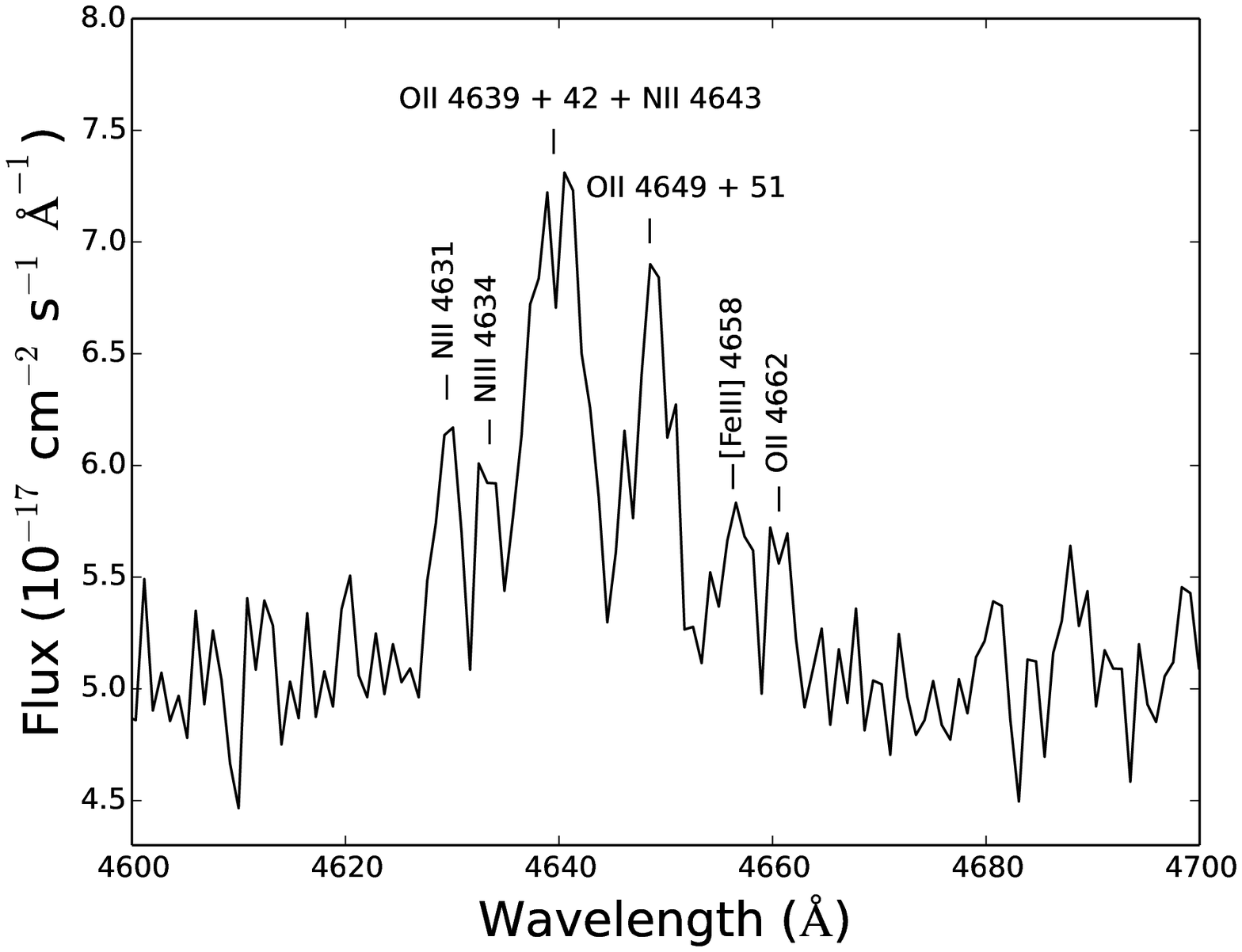} 
   \caption{Section of the OSIRIS spectrum of aperture no. 3 of NGC~7635 --uncorrected for reddening-- showing the recombination lines of multiplet 1 of  {\oii} about 4650 \AA.}
   \label{NGC7635_A3_OII}
  \end{figure}
 %%%%%%%%%%%%%%% 

%%%%%%%%%%%%%%%%%%%%%%%%%%%%%%%%%%%%%%%%%%%%%%%%%%%%%%%%%%%%%%%%%%%%%%%%%%%%%%%%%%%%%%%%%%%%%%%%%
\subsection{Ionic abundances from RLs } \label{rls}
%%%%%%%%%%%%%%%%%%%%%%%%%%%%%%%%%%%%%%%%%%%%%%%%%%%%%%%%%%%%%%%%%%%%%%%%%%%%%%%%%%%%%%%%%%%%%%%%%

We detect several {\hei} emission lines in the spectra of the ring nebulae. Although these lines arise mainly from recombination, those belonging to triplet transitions can be affected by collisional excitation and self-absorption effects. In order to minimize these effects, we determine the He$^+$ abundance of the objects making use of {\sc pyneb} and the recent effective recombination coefficient computations by \citet{porteretal12, porteretal13} for {\hei} lines, where both collisional contribution effects and the optical depth in the triplet lines are included. The final adopted He$^{+}$ abundance is the weighted average of the 
ratios obtained from the individual brightest {\hei} lines measured in each aperture and object and is included in tables~\ref{cond_abun_NGC7635}, \ref{cond_abun_NGC6888}, \ref{cond_abun_G24} and \ref{cond_abun_resto}. The number of {\hei}  lines used in each individual spectrum varies depending on its signal-to-noise ratio. Three lines were used in the case of S~308, five in RCW~52 and RCW~58, between two and five in the apertures of G2.4+1.4 and between 6 and 11 in the cases of NGC~6888 and NGC~7635. We detect {\heii} 4686 \AA\ line in all apertures of G2.4+1.4 due to the high ionisation degree of the nebula, which is ionised by a very hot WO2 star. In the case of NGC~6888, we detect a rather faint {\heii} 4686 \AA\ line in apertures no. 3 to 6. We have determined the He$^{2+}$ abundance using  {\sc pyneb} and the effective recombination coefficient calculated by \citet{storeyhummer95}. The He$^{2+}$ abundances of NGC~6888 and G2.4+1.4 are shown in tables~\ref{cond_abun_NGC6888} and ~\ref{cond_abun_G24}. 

Pure RLs of multiplet 1 of {\oii} have been detected and measured in most apertures of NGC~7635 and NGC~6888 (see figures~\ref{NGC7635_A3_OII} and \ref{NGC6888_A6_OII}, respectively), by the first time in the second object. We have determined the O$^{+2}$ abundance from these lines making use of the effective recombination coefficients  for case B and assuming LS coupling calculated by \citet{storey94} and considering {\elect}(high). As \citet{ruizetal03} pointed out, the relative populations of the levels do not follow local thermodynamical  equilibrium (LTE)  for densities {\elecd} $<$ 10$^4$ {\cmc}. We use the prescriptions of \citet{apeimbertpeimbert05} to calculate the appropriate corrections for the relative strengths between the individual \oii\ lines. 
 
The pure RL \cii\ 4267 \AA\ is detected in several apertures of NGC~7635 and NGC~6888, but only with relatively good signal-to-noise ratio -- error lower than 40\% -- in aperture no. 4 of NGC~7635 (see Figure~\ref{NGC7635_A4_CII}). We detected other {\cii} lines in that aperture, but all of them are produced by fluorescence and not pure recombination. We also report an emission feature in aperture no. 4 of G2.4+1.4 at a wavelength that is consistent 
with {\cii} 4267 \AA\ but the identification is highly dubious (see Section~\ref{total} for a discussion about this issue). We compute the C$^{2+}$ abundance from the measured flux of the \cii\ 4267 \AA\ RL, {\elect}(high), and the \cii\ effective recombination coefficients calculated by \citet{daveyetal00} for case B. The O$^{2+}$ and/or C$^{2+}$ abundances determined from RLs are included in 
tables~\ref{cond_abun_NGC7635}, \ref{cond_abun_NGC6888} and \ref{cond_abun_G24}.

%%%%%%%%%%%%%%% 
  \begin{figure}
   \centering
   \includegraphics[scale=0.45]{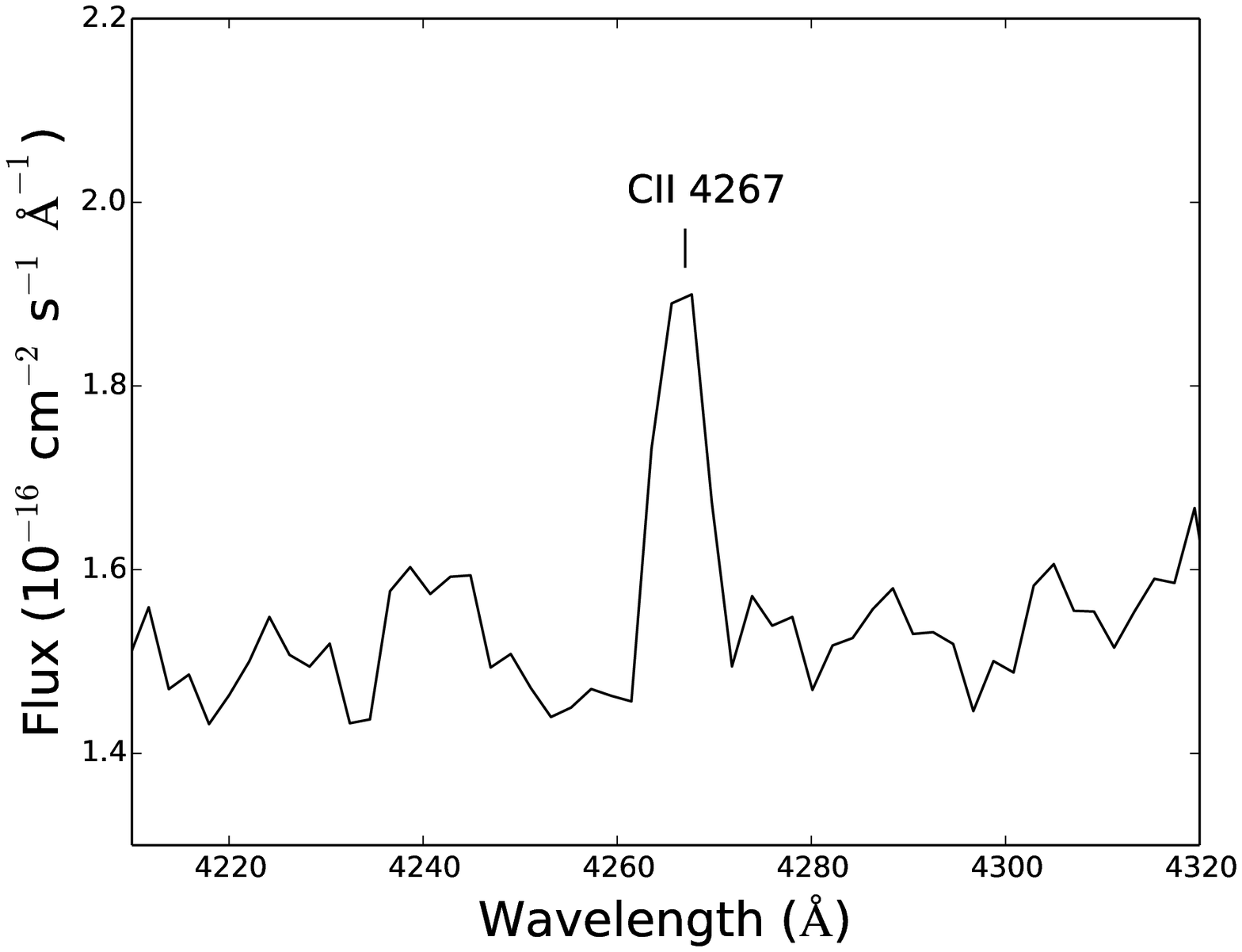} 
      \caption{Section of the OSIRIS spectrum of aperture no. 4 of NGC~7635 --uncorrected for reddening-- showing the recombination line of {\cii} 4267 \AA.}
   \label{NGC7635_A4_CII}
  \end{figure}
 %%%%%%%%%%%%%%% 

%%%%%%%%%%%%%%%%%%%%%%%%%%%%%%%%%%%%%%%%%%%%%%%%%%%%%%%%%%%%%%%%%%%%%%%%%%%%%%%%%%%%%%%%%%%%%%%%%
\subsection{Total abundances } \label{total}
%%%%%%%%%%%%%%%%%%%%%%%%%%%%%%%%%%%%%%%%%%%%%%%%%%%%%%%%%%%%%%%%%%%%%%%%%%%%%%%%%%%%%%%%%%%%%%%%%

To compute the elemental or total abundances, we need to adopt a set of ionisation correction factors (ICFs) to correct for the unseen ionisation stages of each element. 
The O/H ratios are calculated simply adding O$^+$ and O$^{2+}$ abundances in all objects except G2.4+1.4. In this last object we see a very bright {\heii} 4686 \AA\ line, 
indicating a substantial amount of O$^{3+}$. Although the {\heii} line is also detected in some apertures of NGC~6888, its faintness indicates that the amount of O$^{3+}$ is 
negligible (O$^{3+}$/O$^{2+}$ $<$ 0.002) and we will not take its contribution into account. 

To derive He, C, N, Ne, S, Cl and Ar abundances - as well as O ones in the case of G2.4+1.4 - we have used the grid of models build by \citet{delgadoingladaetal14}  and stored in the 3MdB database \citep{http://adsabs.harvard.edu/abs/2015RMxAA..51..103M}. From the models under the reference ``PNe$\_$2014", for each observation of most of the nebulae, we select the models that simultaneously fit the following line ratios: 
[\ion{O}{iii}] 5007/[\ion{O}{ii}] 3727, \ion{He}{i} 5876/H$\beta$ and \ion{He}{ii} 4686/H$\beta$ within 3, 3, and 5 times the observational error respectively. In the case of apertures no. 3, 4 and 6 of NGC~6888 and the single spectrum of RCW~58 we can only reproduce the [\ion{O}{iii}] 5007/[\ion{O}{ii}] 3727 and \ion{He}{ii} 4686/H$\beta$ line ratios. We add to the selection criteria that the models must be radiation-bounded or slightly matter bounded (i.e. the mass of the nebula being greater than 70\% of the radiation-bounded case) and that the nebula is selected as ``realistic" one \citep[the field ``com6" set to one, see][for more details]{delgadoingladaetal14}. The ionizing SED is described by Planck functions as well as T. Rauch models \citep{rauch97}. These models have been run to determine ICFs for Planetary Nebulae (PNe), not for WR stars, but there are no most adapted models in 3MdB, especially that reach very high effective temperatures.
For each observation, the selected models return the needed ICFs for each observed ion and the elemental abundances can be computed. Notice that the process allow us to obtain not a single value for the ICFs, but a set of values that correspond to all the models that reproduce the observational constraints, leading to an estimation of the uncertainties associated to each ICF. These uncertainties are based on an implicit distribution of PNe models, that may not be totally adequate for the objects that we are studying here. We nevertheless think that the method at least give us an idea of the order of magnitude of the range in which we could find the ``real" but inaccessible ICF. To illustrate the way we obtain the ICFs from the 3MdB data, in Figure~\ref{ICFs} we plot the histograms of the ICF values of C$^{++}$, Ne$^{++}$, S$^{+}$ + S$^{++}$, and Ar$^{++}$, for a single observation, namely aperture no. 4 of G2.4+1.4. The green dot indicates the median value and the red line runs over the standard deviation.

%%%%%%%%%%%%%%% 
\begin{figure}
 \centering
 \includegraphics[scale=0.45]{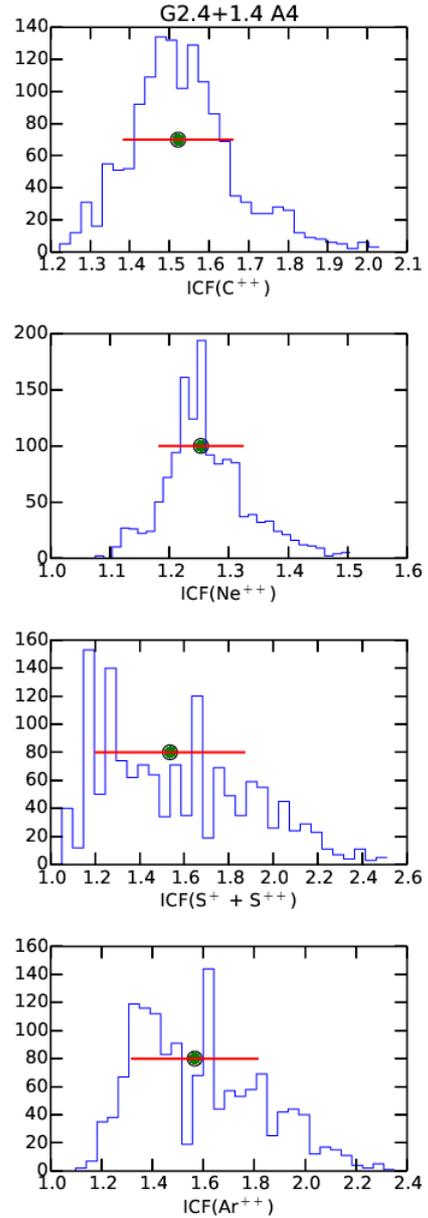} 
 \caption{Histograms showing the frequency of the ICF values of C$^{++}$, Ne$^{++}$, S$^{+}$+S$^{++}$, and Ar$^{++}$ provided by the photoionization models that reproduce the spectrum of aperture no. 
  4 of G2.4+1.4. The green dot indicates the median value and the red line runs over the standard deviation.}
   \label{ICFs}
 \end{figure}
 %%%%%%%%%%%%%%% 

In the case of Fe, we have adopted the ICF scheme computed by \citet{rodriguezrubin05} from observational data (their equation 3). Those authors also derived another ICF using photoionionisation models but it seems to introduce an unexpected trend with the degree of ionisation \citep{delgadoingladaetal09}. As it can be noted in Table~\ref{cond_abun_NGC6888}, we have not calculated total abundances for aperture no. 1 of NGC~6888. This is due no photoionisation models can reproduce the peculiar spectrum of this zone as well as the difference of about 11,000 K between {\elect}({\foiii}) and {\elect}({\fnii}). The characteristics of this particular spectrum will be discussed in Section~\ref{aperture1}. 
%%%%%%%%%%%%%%%%%%%%%%%%%%%%%%%%%%%%%%%%%%%%%%%%%%%%%%%%%%%%%%%%%%%%%%%%%%%%%%%%%%%%%%%%%%%%%%%%%
\subsection{Abundance discrepancy and temperature fluctuations}\label{t2}
%%%%%%%%%%%%%%%%%%%%%%%%%%%%%%%%%%%%%%%%%%%%%%%%%%%%%%%%%%%%%%%%%%%%%%%%%%%%%%%%%%%%%%%%%%%%%%%%%

It is well known that ionic abundances determined from the intensity of CELs and RLs of the same ion do not coincide in nebulae. In the case of {\hii} regions, the O$^{2+}$ abundances derived from RLs are systematically a factor between 1.3 and 3  higher than those computed from CELs \citep[e.g.][]{garciarojasesteban07}. This dichotomy is called the abundance discrepancy problem and its origin is still 
unknown. 

We computed the O$^{2+}$/H$^{+}$ ratios using CELs and RLs in three and five apertures of NGC~7635 and NGC~6888, respectively. As expected, the abundances obtained from RLs are larger than those determined from CELs in all spectra. We can quantify this difference defining the abundance discrepancy factor (hereafter ADF) as the difference between the logarithmic abundances derived from RLs and CELs:

\begin{equation}
\mathrm{ADF(X^{i+}) = log(X^{i+}/H^+)_{RLs} - log(X^{i+}/H^+)_{CELs};}
\end{equation}
where $\mathrm{X}^{i+}$ corresponds to the ionisation state i of element X. The values of the ADF(O$^{2+}$) found in NGC~7635 and NGC~6888 are shown in tables~\ref{cond_abun_NGC7635} and \ref{cond_abun_NGC6888}. The weighted mean of the individual ADF(O$^{2+}$) of the different apertures of NGC~7635 and NGC~6888 are 0.38$\pm$0.12 and 0.50$\pm$0.06 dex, respectively. 
Both values are larger than the typical ADFs reported in the literature for Galactic and extragalactic \hii\ regions \citep[see e.~g.][]{garciarojasesteban07}, 0.26$\pm$0.09 dex,   
but more similar to the mean value found by \citet{estebanetal14} for a sample of star-forming dwarf galaxies, which is 0.43$\pm$0.16 dex. 

There are several hypothesis to explain the origin of the abundance discrepancy problem. The earlier hypothesis is based on the presence of variations in the spatial distribution of {\elect} -- the so-called temperature fluctuations, {\ts} -- in the 
ionised gas of nebulae \citep{torrespeimbertetal80}. In the presence of such fluctuations, the abundances derived from RLs should be more representative of the true ones because abundances relative to hydrogen derived from 
RLs are almost independent on {\elect}, while those determined from CELs are strongly dependent on it. Classical photoionisation models do no predict significant temperature fluctuations but certain  
physical inhomogeneities in nebulae such as the presence of density variations may produce them  \citep{viegasclegg94}. Moreover, departures from a Maxwell-Boltzmann energy distribution of free electrons in nebulae in form of a ``kappa-distribution'' can mimic the effects of temperature fluctuations on the derived abundances \citep{nichollsetal12} and could  explain their ultimate nature. 

\citet{peimbertetal91} explored the effects of shock waves in the spectra of ionised nebulae. They showed that large temperature fluctuations may arise in the presence of shocks with velocities larger than 100 {\kms}. From the observational point of view, there are some indications that the presence of shocks may be related with localized large values of {\ts} and/or ADF in {\hii} regions. \citet{mesadelgadoetal09}, from the analysis of the {\hei} line ratios, found  {\ts} values in the shocked gas of the Herbig-Haro object HH 202-S of the Orion Nebula -- {\ts} = 0.050 -- much larger than in the background nebular gas of the nebula -- {\ts} = 0.016. In addition, \citet{mesadelgadoetal09} find a ADF(O$^{2+}$) = 0.35 for the shocked gas of HH 202-S, in contrast to the value of 0.11 found for the background gas. Moreover, the corresponding {\ts} that these authors obtain assuming that the AD problem is produced by temperature fluctuations is in complete agreement with the {\ts} value obtained from the {\hei} line ratios. HH-204, also in the Orion Nebula, is another Herbig-Haro object showing indications that large {\ts} values may be present in shocked nebular areas. \citet{nunezdiazetal12} find localized 
high values of {\elect} -- about 1000 K larger that the surrounding gas -- in the leading working surface of the shock related to HH-204. As we can see, one possible explanation of the larger ADFs 
and {\ts} in ring nebulae may be the effects of shocks, more important than in {\hii} regions due to their wind-blown nature. Going one step further, we can even speculate that the contribution of unresolved larger shocked areas in more distant starburst-dominated giant {\hii} regions may also be the origin of their high temperature fluctuations and ADFs. In this context, ring nebulae as NGC~7635 and NGC~6888 would be interpreted as Galactic counterparts of those kinds of large-scale shock-contaminated nebulae. 

If we assume the presence of spatial temperature fluctuations as the cause of the abundance discrepancy, then we have to correct abundances obtained from CELs. Under the temperature fluctuations paradigm, the structure of a gaseous nebula is characterized by two parameters: the average temperature, $T_0$, and {\ts}. For each nebula where we measure \oii\ RLs -- NGC~7635 and NGC~6888 -- we can compute the $T_0$ that generates the same O$^{2+}$/H$^+$ ratio from CELs and from RLs. Then, with the corresponding value of {\ts}, we can correct the abundances derived from CELs following the formalism firstly developed by \citet{peimbertcostero69}. The computed ionic and total abundances for {\ts} $>$ 0 are presented in tables~\ref{cond_abun_NGC7635} and \ref{cond_abun_NGC6888}.

%%%%%%%%%%%%%
  \begin{table*}
   \begin{minipage}{180mm}
   \caption{Ionization Correction Factors for NGC~7635 and NGC~6888.}
  \label{icfs1}
    \begin{tabular}{lccccccc}
     \hline
	& & \multicolumn{6}{c}{NGC~7635}  \\
        Ratio & ICF$^{\rm a}$ & A1 & A2 & A3 & A4 & A5 & A6 \\
    \hline
	C/H & C$^{2+}$ & $-$ & 1.39 $\pm$ 0.09 & 1.59 $\pm$ 0.18 & 2.64 $\pm$ 1.41 & $-$ & $-$  \\
	He/H & He$^+$+He$^{2+}$ & 1.53 $\pm$ 0.08 & 1.13 $\pm$ 0.07 & 1.23 $\pm$ 0.12 & 1.58 $\pm$ 0.10 & 1.42 $\pm$ 0.13 & 1.44 $\pm$ 0.11 \\
	N/H & N$^+$ & 1.49 $\pm$ 0.16 & 2.95 $\pm$ 0.37 & 2.32 $\pm$ 0.46 & 1.43 $\pm$ 0.19 & 1.81 $\pm$ 0.22 & 1.76 $\pm$ 0.16 \\
	Ne/H & Ne$^{2+}$ & $-$ & 3.4 $\pm$ 1.0 & 5.1 $\pm$ 1.9 & 18$^{+20}_{-18}$ &  $-$ & $-$  \\
	S/H & S$^+$+S$^{2+}$ & 1.01 $\pm$ 0.01 & 1.04 $\pm$ 0.01 & 1.03 $\pm$ 0.01 & 1.00 $\pm$ 0.01 & 1.01 $\pm$ 0.01 & 1.01 $\pm$ 0.01 \\
	Cl/H & Cl$^{2+}$ & 1.59 $\pm$ 0.39 & 1.19 $\pm$ 0.03 & 1.23 $\pm$ 0.05 & 1.67 $\pm$ 0.62 & 1.36 $\pm$ 0.12 & 1.38 $\pm$ 0.08 \\
	Ar/H & Ar$^{2+}$ & 1.71 $\pm$ 0.11 & 1.20 $\pm$ 0.07 & 1.31 $\pm$ 0.13 & 1.73 $\pm$ 0.16 & 1.52 $\pm$ 0.14 & 1.55 $\pm$ 0.12 \\
	Fe/H & Fe$^{2+}$ & $-$ & 2.2 $\pm$ 1.1 & 2.19 $\pm$ 0.94 & 2.99 $\pm$ 0.39 & $-$ & $-$  \\
    \hline
	& & \multicolumn{5}{c}{NGC~6888}&  \\
	Ratio & ICF$^{\rm a}$& A2 & A3 & A4 & A5 & A6 &  \\
    \hline	
	C/H & C$^{2+}$ & 1.31 $\pm$ 0.61 & 1.54 $\pm$ 0.60 & 1.22 $\pm$ 0.29 & 1.18 $\pm$ 0.16 & 1.22 $\pm$ 0.32 \\
	He/H & He$^+$+He$^{2+}$ & 1.04 $\pm$ 0.12 & 1.08 $\pm$ 0.33 & 1.02 $\pm$ 0.12 & 1.01 $\pm$ 0.02 & 1.02 $\pm$ 0.14 \\
	N/H & N$^+$ & 3.4 $\pm$ 1.4 & 2.12 $\pm$ 0.88 & 5.2 $\pm$ 2.2 & 6.2 $\pm$ 2.1 & 5.0 $\pm$ 2.1 \\
	Ne/H & Ne$^{2+}$ & 1.69 $\pm$ 0.31  & 3.4$^{+9.1}_{-2.4}$ & 1.3$^{+2.0}_{-0.3}$ &  1.26 $\pm$ 0.57 & 1.3$^{+2.3}_{-0.3}$  \\
	S/H & S$^+$+S$^{2+}$ & 1.07 $\pm$ 0.05 & 1.02 $\pm$ 0.03 & 1.16 $\pm$ 0.17 & 1.21 $\pm$ 0.16 & 1.15 $\pm$ 0.14 \\
	Cl/H & Cl$^{2+}$ & 1.26 $\pm$ 0.50 & 1.27 $\pm$ 0.23 & 1.25 $\pm$ 0.24 & 1.25 $\pm$ 0.18 & 1.24 $\pm$ 0.24 \\
	Ar/H & Ar$^{2+}$ & 1.14 $\pm$ 0.10 & 1.15 $\pm$ 0.36 & 1.15 $\pm$ 0.15 & 1.14 $\pm$ 0.12 & 1.14 $\pm$ 0.16 \\
	Fe/H & Fe$^{2+}$ & 3.56 $\pm$ 0.81 & 2.62 $\pm$ 0.89 & 2.47 $\pm$ 0.96 & 2.6 $\pm$ 1.0 & 2.4 $\pm$ 1.0  \\
   \hline
    \end{tabular}
    \begin{description}
      \item[$^{\rm a}$] To be applied as: e.g.~N/H $=$ N$^+$/H$^+ \times$ ICF(N$^+$).       
    \end{description}
   \end{minipage}
  \end{table*}

 %%%%%%%%%%%%%

%%%%%%%%%%%%%
  \begin{table*}
   \begin{minipage}{180mm}
   \caption{Ionization Correction Factors for G2.4+1.4, RCW~52, RCW~58 and S~308.}
  \label{icfs2}
    \begin{tabular}{lccccccccc}
     \hline
	& & \multicolumn{5}{c}{G2.4+1.4} & & & \\
        Ratio & ICF$^{\rm a}$ & A1 & A2 & A3 & A4 & A5 & RCW~52 & RCW~58 & S~308  \\
    \hline
	O/H & O$^+$+O$^{2+}$ & 1.23 $\pm$ 0.03 & 1.32 $\pm$ 0.02 & 1.25 $\pm$ 0.06 & 1.26 $\pm$ 0.07 & 1.23 $\pm$ 0.34 &  $-$ & $-$ & $-$ \\
	C/H & C$^{2+}$ & $-$ & $-$ & $-$ & 1.52 $\pm$ 0.14 &  $-$ &  $-$ & $-$ & $-$ \\
	He/H & He$^+$+He$^{2+}$ & $-$ & $-$ & $-$ & $-$ & $-$ & 1.78 $\pm$ 0.17 & 1.66 $\pm$ 0.80 & 1.03 $\pm$ 0.17 \\
	N/H & N$^+$ & 4.1 $\pm$ 1.6 & 7.05 $\pm$ 0.70 & 3.60 $\pm$ 1.54 & 4.9 $\pm$ 3.0 & 2.3 $\pm$ 1.3 & 0.14 $\pm$ 0.25 & 1.55 $\pm$ 0.47 & 6.3$^{+8.5}_{-6.3}$ \\
	Ne/H & Ne$^{2+}$ & 1.23 $\pm$ 0.03 & 1.34 $\pm$ 0.02 & $-$ & 1.25 $\pm$ 0.07 & 1.20 $\pm$ 0.24 & $-$ & $-$ & 1.2$^{+4.2}_{-1.2}$   \\
	S/H & S$^+$+S$^{2+}$ & $-$ & 1.80 $\pm$ 0.06 & $-$ & 1.54 $\pm$ 0.33 & $-$ & 1.00 $\pm$ 0.01 & 1.01 $\pm$ 0.02 & 1.24 $\pm$ 0.40 \\
	Cl/H & Cl$^{2+}$ & 1.84 $\pm$ 0.09 & 2.05 $\pm$ 0.06 & $-$ & 1.85 $\pm$ 0.27 & $-$ & $-$ & $-$ & $-$ \\
	Ar/H & Ar$^{2+}$ & 1.50 $\pm$ 0.11 & 1.74 $\pm$ 0.05 & 1.43 $\pm$ 0.17 & 1.57 $\pm$ 0.25 & 1.36 $\pm$ 0.19 & 1.96 $\pm$ 0.21 & 1.81 $\pm$ 0.88 & 1.27 $\pm$ 0.31 \\
	Fe/H & Fe$^{2+}$ & $-$ & 2.92 $\pm$ 0.86 & $-$ & $-$ & $-$ & $-$ & 3.9 $\pm$ 1.3 & 3.4$^{+7.7}_{-2.4}$ \\
   \hline
    \end{tabular}
    \begin{description}
      \item[$^{\rm a}$] To be applied as: e.g.~N/H $=$ N$^+$/H$^+ \times$ ICF(N$^+$).       
    \end{description}
   \end{minipage}
  \end{table*}

 %%%%%%%%%%%%%
 
 %%%%%%%%%%%%%%%%%%%%%%%%%%%%%%%%%%%%%%%%%%%%%%%%%%%%%%%%%%%%%%%%%%%%%%%%%%%%%%%%%%%%%%%%%%%%%%%%%
\section{Discussion} \label{sec:discussion}
%%%%%%%%%%%%%%%%%%%%%%%%%%%%%%%%%%%%%%%%%%%%%%%%%%%%%%%%%%%%%%%%%%%%%%%%%%%%%%%%%%%%%%%%%%%%%%%%%

%%%%%%%%%%%%%%%%%%%%%%%%%%%%%%%%%%%%%%%%%%%%%%%%%%%%%%%%%%%%%%%%%%%%%%%%%%%%%%%%%%%%%%%%%%%%%%%%%
\subsection{Abundance patterns } \label{abundances}
%%%%%%%%%%%%%%%%%%%%%%%%%%%%%%%%%%%%%%%%%%%%%%%%%%%%%%%%%%%%%%%%%%%%%%%%%%%%%%%%%%%%%%%%%%%%%%%%%

\subsubsection{NGC~7635}\label{pattern_NGC7635}

Attending to spectroscopical and morphological considerations, the apertures observed in NGC~7635 can be separated into three groups in order to study the abundance pattern in this nebula. The first group is aperture no. 1 that corresponds to one of the brightest knots of S~162 -- a large {\hii} region that encompasses NGC~7635 and is ionised by the same O-type star . The second group corresponds to the brightest high-density ``cometary" knot just at the southwest of the ionising star -- aperture no. 4 -- and the third one would include all the apertures that seem to be part of the main body of the bubble: apertures no. 2, 3, 5 and 6. In fact, all these apertures show lower {\elecd} values and higher ionisation degree than the other two. The mean 
O/H and N/O ratios of the bubble apertures -- considering determinations based on CELs  and {\ts} = 0 -- are 8.43 $\pm$ 0.05 and $-$0.89 $\pm$ 0.08 dex, respectively. \citet{talentdufour79} and \citet{mooreetal02b} obtain values of about 8.49 and $-$1.3 dex for their ``rim" zone, close to our apertures no. 2 and 3. The O/H and N/O ratios we obtain for S~162 are 8.36 $\pm$ 0.17 and $-$0.71 $\pm$ 0.19 dex, respectively, similar to the mean values obtained for the bubble apertures considering the errors. Moreover, the He/H ratios obtained for the bubble and S~162 are also consistent.  As it can be seen in Table~\ref{tab:gradients}, the O/H, N/O and C/H ratios of NGC~7635 are the expected ones for an {\hii} region located at the Galactocentric distance computed by \citet{mesadelgadoesteban10} assuming a Galactocentric distance of the Sun of $R_{\rm 0}$ = 8.0 $\pm$ 0.5 kpc \citep{reid93}. 
Therefore, we can conclude that the bubble of NGC~7635 should be composed by swept-up ambient gas.

The aperture no. 4 of NGC~7635 -- corresponding to the bright dense knot --  shows O/H and N/O ratios of 8.27 $\pm$ 0.03 and $-$0.57 $\pm$ 0.07 dex, respectively, which are slightly different from the bubble values. 
In their detailed study of $HST$ data, \citet{mooreetal02b} indicate that the knots close to the central star of NGC~7635 are the ionised edges of a large mass of neutral material, with strong photoevaporative flows toward the central star. \citet{mooreetal02} demonstrate that the strong density gradients of the knots may produce deviations in the spectrally derived elemental abundances at a level exceeding typical observational errors. Similar results have been found in 
localized high-density areas of the Orion Nebula. \citet{mesadelgadoetal11, mesadelgadoetal12} and \citet{nunezdiazetal12} have found that high-density bright bars, proplyds or Herbig-Haro objects inside the Orion Nebula show somewhat different O/H ratios  
with respect to the almost constant values obtained in the surrounding lower density zones. Those authors conclude that the different O/H ratios may be produced by the presence of dense gas with strong density stratification. In this particular situation, the values of physical conditions and ionic abundances derived using lines of low critical density may be unreliable, giving rise to wrong corrections for collisional de-excitation. 
This effect may be especially important in the calculation of {\elecd}({\fsii}) from the {\fsii} 6716 and 6731 \AA\ lines and the O$^{+}$/H$^{+}$ ratio derived from the {\foii} 3726 and 3729 \AA\ lines. All those lines have 
critical densities about 1-4 $\times$ 10$^3$ {\cmc}, of the order of the values we find for aperture no. 4. Assuming {\elecd}({\fsii}) of about 7000 {\cmc} 
we obtain a total O/H ratio of 8.43 for aperture no. 4, the ratio we obtain for the bubble. The use of this value of density produces a higher O$^{+}$ abundance but leaves O$^{2+}$ virtually unaffected. 
In conclusion, we think that NGC~7635 should be chemically homogeneous and that strong localized density stratification 
may be the reason of the somewhat lower abundances found in aperture no. 4. 
%%%%%%%%%%%%%
  \begin{table*}
   \begin{minipage}{180mm}
     \caption{Abundance comparisons in units of 12+log(X/Y).}
     \label{tab:gradients}
    \begin{tabular}{lcccccccc}
     \hline
     	& $d^{\rm a}$ & $R_{\rm G}^{\rm b}$ & & & & & & \\
       Nebula &  (kpc) & (kpc) & (O/H)$_{\rm grad}^{\rm c}$ & (O/H)$_{\rm obs}$& (N/O)$_{\rm grad}^{\rm d}$ & (N/O)$_{\rm obs}$ & (C/H)$_{\rm grad}^{\rm e}$ & (C/H)$_{\rm obs}$ \\
     \hline
       NGC~7635 - bubble$^{\rm f}$ & 2.4 $\pm$ 0.2 & 9.2 $\pm$ 0.5 &  8.43 $\pm$ 0.07 & 8.43 $\pm$ 0.05 & $-$0.99 $\pm$ 0.11 & $-$0.89 $\pm$ 0.08 & 8.45 $\pm$ 0.14 & 8.57 $\pm$ 0.24$^{\rm g}$ \\
       NGC~7635 - S162$^{\rm h}$ &  &  &  & 8.36 $\pm$ 0.17 & & $-$0.71 $\pm$ 0.19 & & \\
       NGC~6888$^{\rm i}$ & 1.5 $\pm$ 0.5 & 7.8 $\pm$ 0.5 & 8.49 $\pm$ 0.07 & 8.20 $\pm$ 0.03 & $-$0.93 $\pm$ 0.11 & +0.14 $\pm$ 0.08 & 8.59 $\pm$ 0.14 & 8.50: \\
       G2.4+1.4$^{\rm j}$ & 3 $\pm$ 1 & 5 $\pm$ 1 & 8.62 $\pm$ 0.07 & 8.77 $\pm$ 0.09 & $-$0.82 $\pm$ 0.11 & $-$0.27 $\pm$ 0.11 & $-$ & $-$ \\
       RCW~52 & 5.5 & 8.2 & 8.48 $\pm$ 0.07 & 8.91 $\pm$ 0.12 & $-$0.95 $\pm$ 0.11 & $-$0.94 $\pm$ 0.15 & $-$ & $-$ \\
       RCW~58 & 2.7 & 7.4 & 8.51 $\pm$ 0.07 & 8.39 $\pm$ 0.08 & $-$0.92 $\pm$ 0.11 & +0.21$\pm$ 0.16 & $-$ & $-$ \\
       S~308 & 1.5 $\pm$ 0.2 & 9.0 $\pm$ 0.5 & 8.44 $\pm$ 0.07 & 7.98 $\pm$ 0.11 & $-$0.98 $\pm$ 0.11 & $-$0.15 $\pm$ 0.82 & $-$ & $-$ \\ 
            \hline
    \end{tabular}
    \begin{description}
      \item[$^{\rm a}$] References of the assumed distances are given in the text. 
      \item[$^{\rm b}$] Galactocentric distances derived assuming the Sun at 8.0 kpc \citep{reid93}.   
      \item[$^{\rm c}$] Galactic radial gradient determined from O CELs \citep{estebanetal15}.  
      \item[$^{\rm d}$] Galactic radial gradient determined from N and O CELs \citep{carigietal05}. 
      \item[$^{\rm e}$] Galactic radial gradient determined from C RLs \citep{estebanetal05}.
      \item[$^{\rm f}$] The observed values are the average of apertures no. 2, 3, 5 and 6 of NGC~7635.
      \item[$^{\rm g}$] The observed values of C/H are those of aperture no. 4 of NGC~7635.
      \item[$^{\rm h}$] The observed values are those of aperture no. 1 of NGC~7635. 
      \item[$^{\rm i}$] The observed values are the average of apertures no. 3, 4, 5 and 6 of NGC~6888.
      \item[$^{\rm j}$] The observed values are those of aperture no. 2 of G2.4+1.4.
    \end{description}
   \end{minipage}
  \end{table*}
 %%%%%%%%%%%%%

\citet{mesadelgadoesteban10} found C/H ratios between 8.6 to 9.0 dex in several zones of NGC~7635, indicating an abnormally high abundance of this element. However, the comparison of the C/H ratios 
included in Table~\ref{tab:gradients} indicates that the values we determine from our new observations are consistent with those expected from the abundance gradient. We have included the C/H ratio of aperture no. 4 in the table because: a) it is the best determination (the others have uncertainties larger than 40\%) and b) the aforementioned problems associated with density stratification do not affect significantly the abundances determined from RLs. The emissivity of RLs show a much lower dependence on electron density and temperature than CELs. 

 \subsubsection{NGC~6888}\label{pattern_NGC6888}

\citet{kwitter81} and \citet{estebanvilchez92} found N and He overabundances in NGC~6888 and established its stellar ejecta nature. On the other hand, the presence of chemical inhomogeneities in NGC~6888 has been a matter of debate. \citet{estebanvilchez92} found two main kinematical components with different chemical composition, a 
N-rich expanding bubble and an stationary ambient gas with normal abundances. \citet{fernandezmartinetal12} analysed integral field spectroscopy observations in several fields of NGC~6888, finding substantial spatial variations of the chemical composition of the ionised gas. They propose that the nebula is composed by three shells of different origin and chemical composition. Moreover, \citet{stockbarlow14} derived N/O ratios from FIR lines finding spatial variations of that ratio across the nebula. However, by means of photoionisation models, \citet{reyesperezetal15} demonstrate those alleged abundance differences as effects of the ionisation structure within the nebula. 

The comparison of the chemical abundances of NGC~6888 collected in Table~\ref{cond_abun_NGC6888} indicates that apertures no. 3 to 6 show fairly similar physical conditions and ionic abundances. These zones correspond to the bright filaments that conform the 
main body of the nebula (see Figure~\ref{chart_2}). As it has been commented in sections~\ref{conditions} and \ref{total}, aperture no. 1 shows very high values of {\elect}({\foiii}) and covers an external area of rather diffuse low density high-ionised gas \citep[see][]{mooreetal00}, with conditions and ionic abundances very different to the material of the filaments. Aperture no. 2  is also close to the edge of the nebula and, as aperture no. 1, it shows a {\elect}({\foiii}) about 5000 K higher than {\elect}({\fnii}), larger values of O/H and N/H ratios -- but a N/O ratio similar to the filaments -- and lower He abundance. We consider that the spectrum of aperture no. 2 may correspond to the mixing of emission from the filaments and the diffuse high-ionised gas that dominates aperture no. 1. In fact, as it can be seen in Figure 2 of \citet{mooreetal00}, 
the position of our aperture no. 2 corresponds to a tiny filament embedded in the {\foiii}-dominated tenuous gas. 
The mean values of the O/H, He/H and N/O ratios of apertures from no. 3 to no. 6 -- the filaments -- are 8.20 $\pm$0.03, 11.22 $\pm$0.01 and +0.14 $\pm$ 0.08 dex respectively (see also Table~\ref{tab:gradients})  confirming the important N and He enrichment -- as well as an O deficiency -- that NGC~6888 shows, in agreement with previous studies. The O/H, He/H, and N/O ratios obtained by \citet{mesadelgadoetal14} for a zone almost coincident with our aperture no. 6, are in very good agreement with our results. Our determinations of the C/H ratio in this object have large uncertainties but seems to be fairly similar to the C/H ratio expected from the abundance gradient at the Galactocentric distance assumed by \citet[][see Table~\ref{tab:gradients}]{mesadelgadoetal14}. Therefore, there is not 
compelling reason to consider anomalous C/H ratios in this nebula as the previous results by \citet{mesadelgadoetal14} suggested. This result would lead to the interesting result that the ON-cycle of the CNO burning has been more effective than the CN-cycle in the interior of the massive progenitor star of NGC~6888. This suspicion is also supported by the recent results obtained by \citet{vamvatira-nakouetal16} from {\it Herschel} observations of M1-67, also a stellar ejecta nebula associated to a massive WR star. These authors determine the C/O from the photodissociation region (PDR) line fluxes finding it consistent with the solar C/O ratio within the errors. 

\subsubsection{G2.4+1.4}\label{pattern_G2414}

Due to the faintness of G2.4+1.4, its spectra are of comparatively much lower signal-to-noise ratio than those of NGC~7635 and NGC~6888 and abundance determinations are less precise for this object. The 12+log(O/H) shows  
a rather wide range of values from 8.27 to 8.79 in the different apertures. We must recall that the O abundance of this nebula is derived assuming an ICF(O), so it may be affected by biases difficult to estimate. On the other hand, the O abundances of apertures no. 1, 2 and 5 are very similar, but those of apertures no. 3 and 4 show 
lower values. It is interesting to note that these two last apertures show precisely the highest values of {\elect} and the lowest total abundances determined from CELs. The combination of 
both facts suggests that {\elect} may be somewhat overestimated in these zones or affected by localized shock heating in a manner similar to that detected in aperture no. 1 of NGC~6888 (see Section~\ref{aperture1}). A significant result is 
that although the O abundance shows some variation from one aperture to another, the N/O ratio is fairly constant. In Table~\ref{tab:gradients} we compare the O/H and N/O ratios of 
aperture no. 2 -- the spectrum with the best signal-to-noise ratio of this object -- with the values expected from the Galactic abundance gradients. We have assumed the photometric distance of 3 $\pm$ 1 kpc 
determined by \citet{dopitaetal90}. The O/H ratio is somewhat high in  G2.4+1.4, but seems to be consistent with the value of the O abundance expected at $R_{\rm G}$ $\sim$ 5 kpc  within the errors . 
The log(N/O) of $-$0.27 is rather high and indicates some N-enrichment in the nebular gas. In the case of the He abundance, all the apertures except no. 1 give a mean 12+log(He/H) of about 11.02. Since there is not evidence of a radial abundance gradient of He in the Milky Way, all we can say is that the mean He/H ratio is consistent with that expected for the solar vicinity. \citet{estebanetal92} derived abundances in several zones of G2.4+1.4 and found an O abundance of 8.45 and a N/O ratio of $-$0.49, values that agree with our determinations within the uncertainties. However, \citet{estebanetal92} also found a 12+log(He/H) of about 11.35, which is not confirmed with our deeper spectra. Summarizing, our results do not provide definitive evidence of chemical inhomogeneities in G2.4+1.4, only indications of a N enrichment. 

As it has been commented in section~\ref{rls}, the \cii\ 4267 \AA\ line is only detected in aperture no. 4 of G2.4+1.4, and its intensity is abnormally high. The corresponding 12+log(C/H) results to be 10.21, an excessively high value even taking into account the relatively small Galactocentric distance of the nebula and the expected value of the C/H ratio extrapolating the radial C abundance gradient determined by \citet{estebanetal05}. We have inspected the spectra carefully finding no clear explanation for this oddity. Perhaps an artifact on the detector or an unknown emission feature is mimicking an emission line just at the wavelength of \cii\ 4267 \AA. We have to wait for an independent observational confirmation to ascertain whether this is a real feature or not. 

\subsubsection{RCW~52}\label{pattern_RCW52}

This work presents the first abundance determinations of RCW~52. As it can be seen in Table~\ref{tab:gradients}, the O/H ratio of 8.91$\pm$0.12  is very high with respect to that expected for the Galactocentric distance of the nebula although its N/O is normal, so the object does not show traces of N enrichment. The He abundance of 11.13$\pm$0.05 and also seems somewhat high. However, the ICF(He$^+$) for this nebula is the largest one of the sample, so the exact value of He/H is highly uncertain. We have assumed the photometric distance of 5.5 kpc determined by \citet{reed93} for this object. 

\subsubsection{RCW~58}\label{pattern_RCW58}

There are two previous determinations of the chemical abundances of RCW~58 \citep{kwitter84, rosamathis90}, but both are based on an assumed value of {\elect}. Therefore, our set of abundances is the first one that has been determined using  a direct determination of {\elect}, {\elect}({\fnii}) in this case. The comparison with the abundance gradients shown in Table~\ref{tab:gradients} confirms that the nebula is strongly N enriched. Moreover, the 12+log(He/H) is 11.46$\pm$0.21, a very high value but affected by large uncertainties in the ICF(He). Although the O/H ratio is consistent with the abundance gradient within the errors, we cannot rule out some slight O deficiency in the object. We have assumed the photometric distance of 2.7 kpc determined by \citet{heraldetal01} for this object.

\subsubsection{S~308}\label{pattern_S308}

The chemical abundances of S~308 were determined by \citet{kwitter84, rosamathis90} and \citet{estebanetal92} and all of them found high N/O and He/H ratios in the nebula. Only \citet{rosamathis90} and \citet{estebanetal92} determined {\elect}({\foiii}) in this high-ionised object. In our case, we also determine consistent values of {\elect}({\fnii}) and {\elect}({\fsiii}). The comparison with the abundance gradients 
shown in Table~\ref{tab:gradients} indicates that S~308 seems to be O-deficient in a higher degree than the N-rich filaments of NGC~6888. It also shows high N/H and He/H ratios. Considering that the O, N, Ne, S and Ar abundances -- all elemental abundances determined from CELs -- of S~308 are lower with respect to the rest of the objects, as well 
as its very high values of {\elect}, we cannot rule out the possibility that {\elect} may be overestimated due to an extra-heating of the gas due to shocks. We have assumed the distance of 1.5 kpc proposed by \citet{chuetal03} for S~308.

%%%%%%%%%%%%%%%%%%%%%%%%%%%%%%%%%%%%%%%%%%%%%%%%%%%%%%%%%%%%%%%%%%%%%%%%%%%%%%%%%%%%%%%%%%%%%%%%%
\subsection{Shock excitation in aperture no. 1 of NGC~6888 } \label{aperture1}
%%%%%%%%%%%%%%%%%%%%%%%%%%%%%%%%%%%%%%%%%%%%%%%%%%%%%%%%%%%%%%%%%%%%%%%%%%%%%%%%%%%%%%%%%%%%%%%%%

%%%%%%%%%%%%%%% 
  \begin{figure}
   \centering
   \includegraphics[scale=0.45]{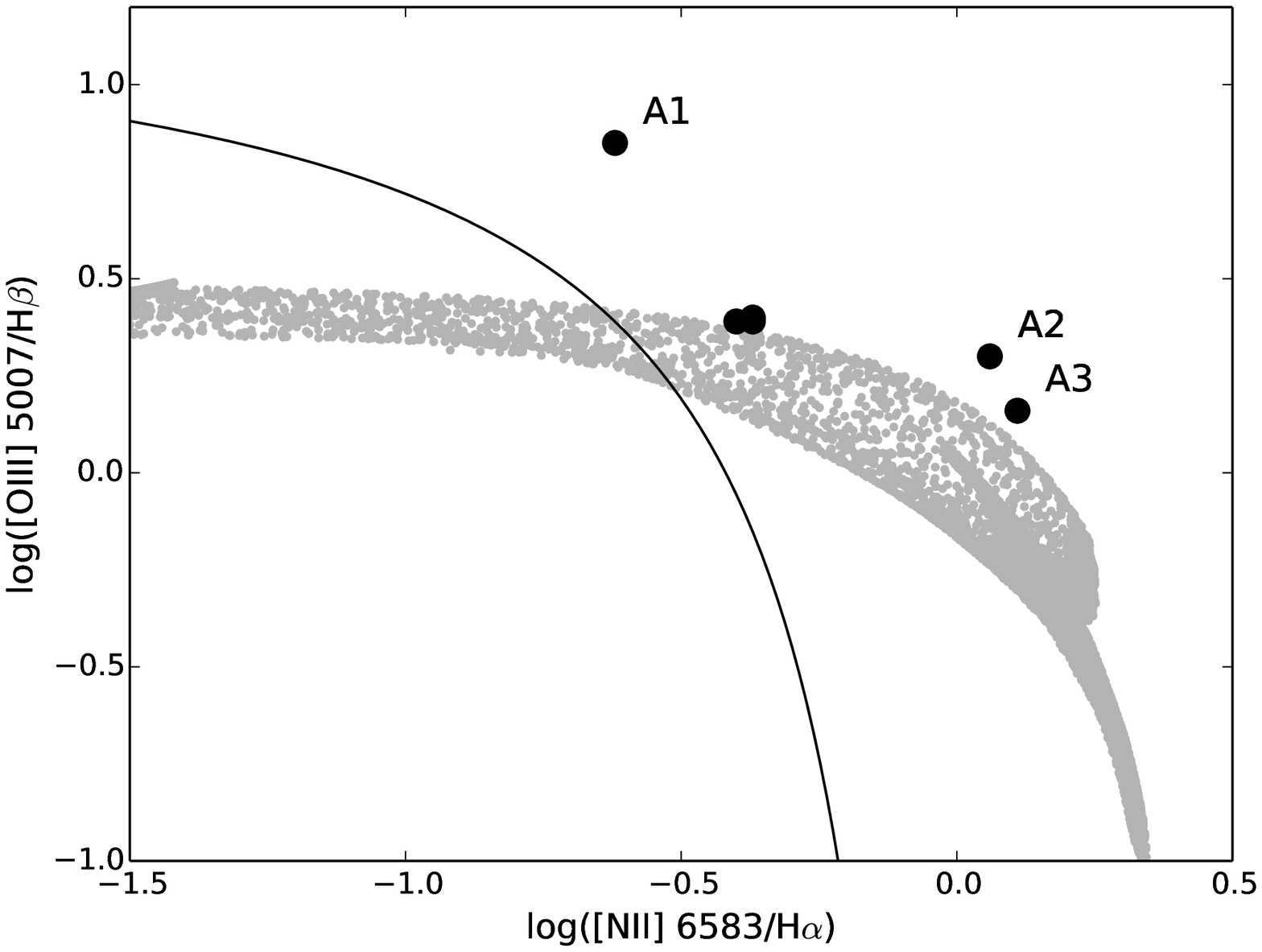} 
      \caption{Diagnostic diagram of {\foiii} 5007/H$\beta$ vs. {\fnii} 6583/H$\alpha$. The solid line indicates the separation between photoionisation models 
      of star-forming regions and AGNs as defined by \citet{kauffmannetal03} for their assumed metallicity. The grey points correspond to the 
      photoionization model of NGC~6888 \citet{reyesperezetal15}. They do not fit the separation line because of the larger N/O ratio of the nebula. Black circles 
      correspond to the spectra of the apertures of NGC~6888. The size of their errorbars are of the order or smaller than the circles. The three apertures at the north of 
      the nebula, A1, A2 and A3 are indicated. }
   \label{BPT}
  \end{figure}
 %%%%%%%%%%%%%%% 

As it has been commented in previous sections, aperture no. 1 of NGC~6888 shows a very high value of {\elect}({\foiii}) and we cannot derive its total abundances. This is because we can not find a photoionization model that reproduces its spectrum and, therefore, we lack of an appropriate set of ICFs. 
NGC~6888 shows X-ray emission that is especially bright in the northern part of the nebula \citep[e.g.][]{wriggeetal94, toalaetal14}, a feature that should be produced by collision between the WR wind and a preexisting circumstellar shell \citep[e.g.][]{garciaseguraetal96}. The strong difference between {\elect}({\foiii}) and {\elect}({\fnii}) we find -- in the sense that {\elect}({\foiii}) is higher than {\elect}({\fnii}) --  is a common feature in the spectra of supernova remnants 
\citep[e.g.][]{fesenkirshner82, fesenetal82,sankritetal98}. \citet{fernandezmartinetal12} indicate the possible presence of shocks in the northern area of NGC~6888. To illustrate that possibility in their  Zone B, these authors compare their data with the library of fast radiative shock models of \citet{allenetal08}. However, a drawback of the comparison made by \citet{fernandezmartinetal12} -- especially important in the upper panel of their Figure 14 -- is that the N/O ratio of NGC~6888 is almost one order of magnitude larger than solar, and the shock models they used assume a twice solar abundance. On the other hand, \citet{reyesperezetal15} shows that photoionisation is the main excitation mechanism of NGC~6888 using their tailored photoionisation models and observational points in a BPT \citep{baldwinetal81} diagram (their Figure 7). 
In our Figure~\ref{BPT}, we show an adaptation of Figure 7 of \citet{reyesperezetal15} that represents {\foiii} 5007/{\hb} versus {\fnii} 6584/{\ha} line intensity ratios. The solid line indicates the separation between 
photoionisation models of star-forming regions and active galactic nuclei (AGNs) as defined by equation 1 of \citet{kauffmannetal03}. The grey points illustrate the individual spaxels of the 2D images obtained by 
\citet{reyesperezetal15} from their photoionisation model of NGC~6888 that assumes chemical homogeneity in the nebula as well as abundances very similar to ours. As we can see, the points of the models spread in a relatively wide region of the diagram, indicating the rather different local ionisation conditions 
across the nebula. The black circles correspond to the line ratios measured in all apertures of NGC~6888. It is remarkable that the line defined by the observational points of apertures no. 2 to 6 is roughly parallel to the general shape of the region of the photoionisation model. This indicates that the bulk of the gas of apertures no. 2 to 6 is photoionised. On the other hand, the point corresponding to the spectrum of aperture no. 1 is clearly above the photoionisation model and apart from the rest of the observational points. This indicates that its spectrum is not dominated by photoionisation, conversely, its position is consistent of having an important contribution of shock excitation \citep[see also similar diagrams by][]{allenetal08}. Aperture no. 2 is also slightly above the photoionisation area suggesting some contamination of shock excitation. 

%%%%%%%%%%%%%%%%%%%%%%%%%%%%%%%%%%%%%%%%%%%%%%%%%%%%%%%%%%%%%%%%%%%%%%%%%%%%%%%%%%%%%%%%%%%%%%%%%
\subsection{Comparison with stellar evolution models} \label{models}
%%%%%%%%%%%%%%%%%%%%%%%%%%%%%%%%%%%%%%%%%%%%%%%%%%%%%%%%%%%%%%%%%%%%%%%%%%%%%%%%%%%%%%%%%%%%%%%%%

\citet{estebanvilchez92} and \citet{estebanetal92} compared the abundance pattern of Galactic ejecta ring nebulae with the surface abundances predicted by the evolutionary models of massive stars of  \cite{maeder90}. 
They found that the He, N and O abundances of the nebulae are consistent with the expected nucleosynthesis for stars with initial masses between 25 and 40 M$_\odot$ at a given moment of 
the red supergiant (RSG) phase, prior to the WR stage. A similar initial stellar mass range was obtained by \citet{garnettchu94} for the WN4 star Br~13 in the 
Large Magellanic Cloud, from abundances of its associated ejecta nebula. \citet{mesadelgadoetal14} compared the He, C, N and O abundances they determine for NGC~6888 with the predictions of the stellar evolution models of \citet{ekstrometal12} and \citet{georgyetal12}. Their conclusions are also consistent with those obtained in previous works, the ionising star of NGC~6888 should have an 
initial mass between 25 and 40 M$_\odot$. A similar comparison has been made by \citet{vamvatira-nakouetal16} in the case of M~1-67, finding a good fit for a central star of 32 M$_\odot$.  

 %%%%%%%%%%%%%
  \begin{table*}
   \begin{minipage}{180mm}
     \caption{Some relevant data of central stars and associated ejecta ring nebulae.}
     \label{tab:stars}
    \begin{tabular}{lcccccc}
     \hline
     	& & & $v_{\rm rot}^{\rm b}$ & $M_*$ & & Dyn. Age \\
       Nebula &  Star & Spec. Type$^{\rm a}$ & (\kms) & ($M_\odot$) & Binarity$^{\rm a}$ & ($\times$10$^4$ yr) \\
     \hline
       NGC~6888 &  WR136 & WN6 & 37 & 15$^{\rm c}$ & ? & 2$-$4$^{\rm d}$\\
       G2.4+1.4 & WR102 & WO2 & $-$ & $\sim$9$^{\rm e}$ & $-$ & 8.8$^{\rm f}$, 34$^{\rm g}$\\
       RCW~58 & WR40 & WN8 & 113 & 18$-$34$^{\rm h}$ & $-$ & 5.1$^{\rm f}$\\
       S~308 & WR6 & WN4 & 36 & 19$^{\rm c}$ & ? & 13.5$^{\rm f}$\\
            \hline
    \end{tabular}
    \begin{description}
      \item[$^{\rm a}$]  Data from the Galactic Wolf Rayet Catalogue v1.15 (http://www.pacrowther.staff.shef.ac.uk/WRcat/).
      \item[$^{\rm b}$]  \citet{grafeneretal12}.
      \item[$^{\rm c}$] Present-day stellar mass determined by \citet{hamannetal06}.  
      \item[$^{\rm d}$] \citet{freyeretal06}.
      \item[$^{\rm e}$] Present-day stellar mass determined by \citet{tramperetal15}.  
      \item[$^{\rm f}$] \citet{gruendletal00}.
      \item[$^{\rm g}$] \citet{gosachinskijlozinskaya02}
      \item[$^{\rm h}$] Present-day stellar mass determined by \citet{heraldetal01}.
    \end{description}
   \end{minipage}
  \end{table*}
 %%%%%%%%%%%%%
 
 %%%%%%%%%%%%%%% 
  \begin{figure*}
   \centering
   \includegraphics[width=177mm]{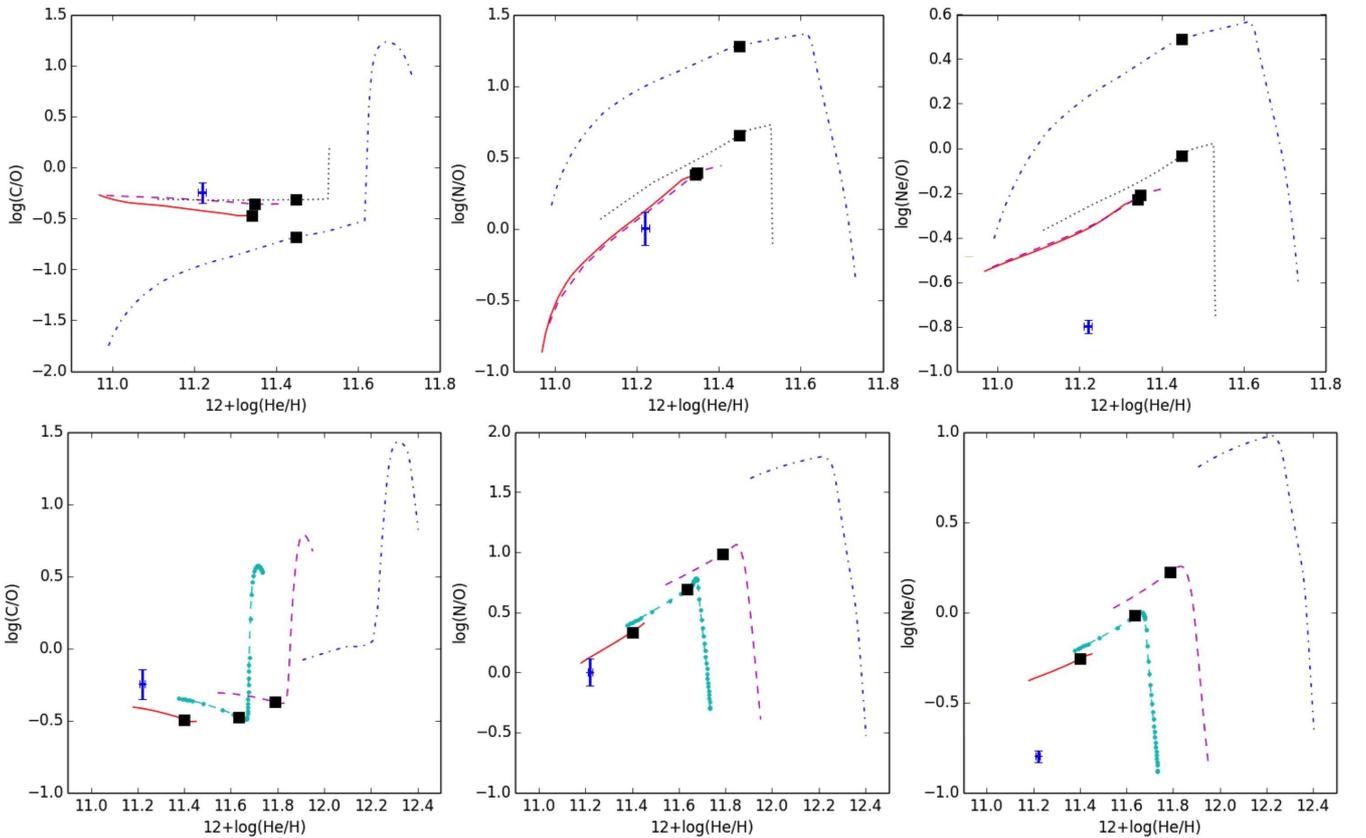} 
   \caption{Abundance ratios of C/O, N/O and Ne/O as a function of the He abundance of the stellar material ejected by stars of different initial masses from non-rotation stellar evolution models (upper panels) and 
   rotation ones (lower panels). The models are by \citet{ekstrometal12}. The blue point corresponds to our observed abundances for NGC~6888. The stellar masses represented are: 25 $M_{\odot}$ (red solid line), 32 $M_{\odot}$ (cyan dashed-big dotted line, only represented for rotational models), 40 $M_{\odot}$ (magenta dashed 
   line), 50 $M_{\odot}$ (black dotted line, only represented for non-rotational models) and 60 $M_{\odot}$ (blue dotted-dashed line). The black square indicates the onset of the WR phase for each model.}
   \label{models_NGC6888}
  \end{figure*}
 %%%%%%%%%%%%%%% 

We have applied the same procedure followed by \citet{mesadelgadoetal14} to our sample of ejecta ring nebulae in order to constrain the mass of their progenitor stars. 
We have used the set of mean He, C, N, O and Ne abundances determined for each object and the stellar evolution models of \citet{ekstrometal12} and \citet{georgyetal12}. We have considered rotational and 
non-rotational models 
for stars with initial masses of 25, 32, 40, 50 and 60 $M_{\odot}$. For the calculations, we have assumed that the stellar surface abundances given by the models are representative of the chemical composition of the stellar wind. The surface abundances of H, He, C, N, O, and Ne given by the models were weighted by the mass-loss rate and then integrated over time using the composite trapezoidal rule. In all the models, the initial integration point was set at the moment in which the massive star reaches the minimum effective temperature, $T_{eff}$, after finishing the main sequence (MS) stage. This point exactly coincides with the onset of enhanced mass loss at the beginning of the RSG phase (or luminous blue variable phase for the most massive stars) and, therefore, our approximation does not include material expelled during the MS stage. As \cite{toalaarthur11} argued, the contribution of the material ejected during the MS can be ignored due to it is spread throughout a much more extended bubble with a size of several tens of parsecs, while the RSG and WR wind material would form a ring nebula in the immediate surroundings around the central star. To properly compare the model predictions with the observed abundances, the integration results for He, C, N, O, and Ne relative to H and/or O were converted from fractional mass to number units through the atomic weight of the elements. 
We used the He/H ratios to explore the variation of the He abundance because they are calculated from the intensity of RLs in both numerator and denominator. We also used the O abundance determined from RLs for determining the C/O ratio in the case of NGC~6888. However, since the N and Ne abundances are always determined from CELs, we used their ratios with respect to the O abundances also derived from CELs 
for all the objects. Considering those particular abundance ratios, we avoid abundance discrepancy effects in the comparisons. In figures~\ref{models_NGC6888} to \ref{models_RCW58}, we represent the assumed  
mean abundance ratios determined for NGC~6888, G2.4+1.4, S~308 and RCW~58 with respect to their He abundance and compare with the predicted abundance ratios of the stellar material ejected along time  by the stellar models for 25, 32, 40, 50 and 60 $M_{\odot}$ stars. 

In Table~\ref{tab:stars} we compile some relevant data of the central star and the nebula of our stellar ejecta objects. We include the identifier of the star, its spectra type, rotational velocity and estimation of its present-day mass. In addition, we also include information about the binarity of the central star. In all the cases, we do not have clear indications of binarity and therefore we can consider them as single stars. 
Finally, we include the dynamical age estimated for the nebulae. 

\subsubsection{NGC~6888}\label{comp_NGC6888}

Figure~\ref{models_NGC6888} shows the comparison for NGC~6888. This is the only object of our sample of WR ring nebulae for which we have derived the C abundance, and can compare the C/O ratios in addition to N/O and Ne/O ones. In Figure~\ref{models_NGC6888} we can see the important differences between the non-rotational models (upper panels) and rotational ones (lower panels). Though non-rotational models 
do not show significant differences for the nucleosynthesis produced by stars of initial masses between 25 and 40 $M_{\odot}$, the situation is completely different in the rotational case. One of the most important effects of the stellar rotation is that the WR phase starts much earlier than what is predicted by the non-rotational models. In figures~\ref{models_NGC6888} to \ref{models_RCW58} the onset of the WR phase is indicated by squares. In all cases, this was set following the definitions of stellar types of \citet{Eldridgeetal08} and the considerations discussed by \citet{georgyetal12}. 
In the figures, we can see that rotational models of the 60$M_\odot$ star do not show a black square, this is because the models predict that the WR phase initiates during the MS stage in these stars.

An inspection of Figure~\ref{models_NGC6888} indicates that the C/O and N/O ratios of NGC~6888 are consistent with non-rotational models of 25 to 40 $M_{\odot}$ and only with the rotational model of 25 $M_{\odot}$. In both cases, the fit between the observational data and models is produced for surface stellar abundances in a moment between the beginning of the SGR and before the onset of the WR phase. 
It must be noted that rotational models are calculated assuming a very large initial rotation velocity of about 300 km s$^{-1}$, much larger than the present-day value of 37~\kms\ calculated for the central star of NGC~6888 (see Table~\ref{tab:stars} and references therein). Therefore, the non-rotational models seem to be more appropriate for this particular object. In fact, the 25 to 40 $M_{\odot}$ range we find with 
the non-rotational models agrees with the initial mass of about 32 $M_{\odot}$ estimated by \citet{georgyetal13} from the stellar parameters of WR136 given by \citet{hamannetal06} and the HR diagram tracks calculated by \citet{ekstrometal12}. Another indication is the comparison between the dynamical age of the nebula with respect to the onset of the WR stage and the position of the observational points in the diagrams. 
NGC~6888 has a rather small dynamical age (2-4 $\times$ 10$^4$ yr see Table~\ref{tab:stars}) indicating that the enriched material of the nebula was ejected close to the onset of the WR phase. This situation is more consistent with the results obtained in the non-rotational case. \citet{mesadelgadoetal14} determined the C/H ratio for NGC~6888 but they do not fit it with any stellar evolution model (see their figures 4 and 5). Fortunately, our new determination of the C abundance from  OSIRIS spectra does not show that disagreement. However, a problem with Ne/O ratio still persists. The Ne/O ratio we obtain -- as well as the previous determination by \citet{mesadelgadoetal14} -- is between 0.3 and 0.5 dex lower than the predictions of the stellar evolution models. This problem will be discussed in Section~\ref{low_neo}. 

Note that we use the O/H ratio determined from CELs for calculating the observed N/O and Ne/O ratios and the O/H determined from RLs for C/O in Figure~\ref{models_NGC6888}. The agreement between models and observations would be worse if we use O/H from CELs for C/O and O/H from RLs for N/O and Ne/O.  In this case the observation point would be displaced about +0.5 dex or $-$0.5 dex in ordinates, respectively. This is consistent with the idea that the AD problem 
is related to the kind of line that we are observing and that abundance ratios using the same kind of lines are less affected by the AD. On the other hand, the comparison between O/H ratios determined from CELs and RLs with the evolution of the O/H predicted by the models does not give an answer about which kind of lines gives the correct abundances. The comparison depends basically on the O/H ratio assumed by the models, which is 12+log(O/H) = 8.66. The observed O/H ratios observed from CELs are always below the 25, 32 and 40 $M_{\odot}$ non-rotational models before the onset of the WR phase. Observed O/H ratios determined from RLs seem to be more consistent with the models.

\subsubsection{G2.4+1.4}\label{comp_G2414}

  %%%%%%%%%%%%%%% 
  \begin{figure}
   \centering
   \includegraphics[width=85mm]{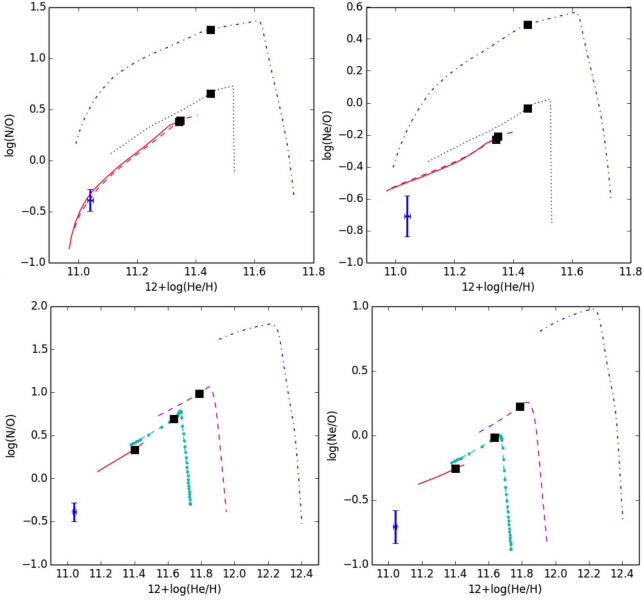} 
   \caption{Same as Figure~\ref{models_NGC6888} for G2.4+1.4 and only including N/O and Ne/O ratios as a function of the He abundance.}
   \label{models_G2414}
  \end{figure}
 %%%%%%%%%%%%%%% 

In the case of G2.4+1.4 and S~308, we can not derive C abundances and we only compare the N/O, Ne/O and He/H ratios with the predictions of stellar evolution models. Figure~\ref{models_G2414} 
shows the comparison between data and models for G2.4+1.4. We can see that the rotational models fail to reproduce the observations, but the non-rotational models of 25 to 40 $M_{\odot}$ fit 
rather well the observed N/O vs. He/H ratios. \citet{tramperetal15} compare stellar parameters of Galactic WO stars and the HR diagram tracks provided by \citet{ekstrometal12} and \citet{georgyetal12}, finding that the likely progenitors of WO stars have initial masses of about 40-60 $M_{\odot}$. On the other hand, \citet{grohetal13} estimate that the progenitors of WO1-3 stars would be stars of initial masses higher than 40 $M_{\odot}$, in the case of assuming the non-rotational models by \citet{ekstrometal12} and \cite{georgyetal12} and equal or higher than 32 $M_{\odot}$ assuming the rotational models. The rotational velocity of WR102 is actually unknown. Although \citet{sanderetal12} report on a very high rotational velocity, 
\citet{tramperetal15} do not find any indication of rapid rotation in WR102. As we can see, only an initial mass of about 40 $M_{\odot}$ for WR102 would reconcile the nebular and stellar results. 
There are two estimations of the dynamical age of the nebula in the literature \citep{gruendletal00, gosachinskijlozinskaya02} and they differ almost a factor four (see Table~\ref{tab:stars}). The comparison of the N/O ratios shown in 
Figure~\ref{models_G2414} indicates that the stellar material should be ejected closer to the beginning of the RSG phase than to the onset of the WR stage. This would better agree with the higher value of the dynamical age of the nebula. 
\citet{tramperetal15} estimates that the central WO2 star of G2.4+1.4 (WR102) will explode as a SNIc in less than 2000 yr. Therefore, G2.4+1.4 stands as an invaluable example of the pre-existent circumstellar medium in which the shock front of a SN explosion will develop. 
 
\subsubsection{S~308}\label{comp_S308}

 %%%%%%%%%%%%%%% 
  \begin{figure}
   \centering
   \includegraphics[width=85mm]{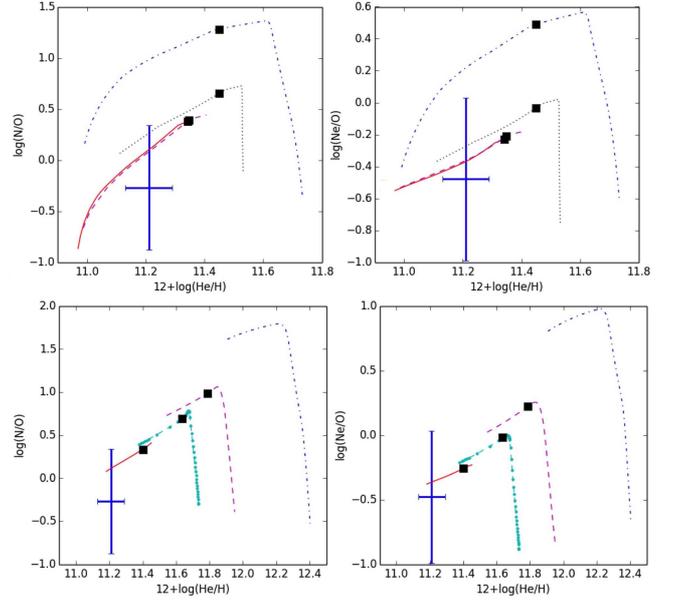} 
   \caption{Same as Figure~\ref{models_G2414} for S~308.}
   \label{models_S308}
  \end{figure}
 %%%%%%%%%%%%%%% 
 
Figure~\ref{models_S308} compares the N/O, Ne/O and He/H ratios of S~308 with the predictions of stellar evolution models. The big error bars of the N/O and Ne/O ratios are due to the large uncertainties 
of the ICFs of N and Ne for this object. In fact, although its ICF(Ne) is very uncertain, we have assumed a nominal error of $\pm$0.30 dex for the Ne/O ratio. As we can see in Figure~\ref{models_S308}, both observed abundance ratios of S~308 (N/O and Ne/O vs. He/H) are consistent with the surface stellar abundances of non-rotational models of stars of initial masses between 25 and 50 $M_{\odot}$. In the case of rotational models, the fit is only possible with stars of 25 $M_{\odot}$. The present-day rotational value of WR6, the central star of S~308, is 36~\kms\ (see Table~\ref{tab:stars} and references therein) very far from the 300~\kms\ assumed in the rotational stellar evolution models. The dynamical age of the nebula is about 13.5 $\times$ 10$^4$ yr \citep{gruendletal00}, so the ejection of the stellar material should be produced 
not far before the onset of the WR phase. 

\subsubsection{RCW~58}\label{comp_RCW58}
 
The low ionisation degree of RCW~58 does not permit to detect {\fneiii} lines and to derive the Ne abundance. We can only compare the observed N/O and He/H ratios with the predictions of the stellar evolution models. In Figure~\ref{models_RCW58} we present that comparison. The large uncertainty of the ICF(He) implies that the error bar of the He/H ratio is very large. The observational point does not fit the surface 
stellar abundances predicted by any model except for the non-rotational models of a 50 $M_{\odot}$ star. However, the fit occurs at a point after the onset of the WR stage, which is not a reasonable solution because contradicts the current picture of the origin of stellar ejecta ring nebulae. If we assume that the He/H ratio of RCW~58 is overestimated by our ICF(He) and that our N/O is correct -- this is directly based in the observations -- the observational data 
may be consistent with the non-rotational models of 25 to 40 $M_{\odot}$ and the rotational ones of 25 $M_{\odot}$.

   %%%%%%%%%%%%%%% 
  \begin{figure}
   \centering
   \includegraphics[width=85mm]{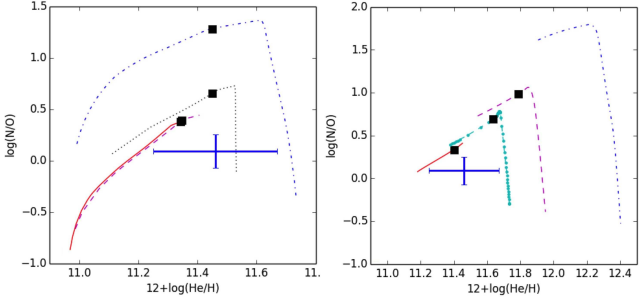} 
   \caption{Same as Figure~\ref{models_NGC6888} for RCW~58 and only including N/O ratios as a function of the He abundance. Left panel: non-rotation stellar evolution models; right panel: rotation models.}
   \label{models_RCW58}
  \end{figure}
 %%%%%%%%%%%%%%% 
 
 \subsubsection{The low Ne/O ratios}\label{low_neo}

Figures~\ref{models_NGC6888} and \ref{models_G2414} indicate that the Ne/O ratios of NGC~6888 and G2.4+1.4 are below the model predictions for both, non-rotational and rotational models. 
This discrepancy was also reported in NGC~6888 by \citet{mesadelgadoetal14}. They find a 12+log(Ne/H) of 7.51$\pm$0.20, almost coincident with our mean value of 7.52$\pm$0.01, which are 
about 0.5 dex lower than the Ne abundance of the Orion Nebula, 12+log(Ne/H) =  $8.01\pm0.01$ \citep{rubinetal11}. Although Ne abundances are based on the assumption of an ICF(Ne), the large 
difference can not be explained solely by systematical uncertainties introduced by this parameter.  To understand such underabundance of Ne, \citet{mesadelgadoetal14} invoked the action of the NeNa cycle \citep[e.g.][]{arnouldmowlavi93}. This cycle destroys Ne nuclei in the H-burning zone at temperatures of about $3.5\times10^7$ K, values that can be reached in massive star cores. Enhancement factors of 6-10 have been predicted by nucleosynthesis models \citep{woosleyetal95, chieffietal98}. Large Na abundances have been observed in the more massive yellow supergiants. \citet{denissenkov05} has proposed that rotational mixing of Na between the convective core and the radiative envelope in the MS progenitors 
can account for the overabundances observed in those objects. Very recently, \citet{huenemoerderetal15} have reported a significantly enhanced Na abundance in the X-ray spectrum of 
the wind of WR6, the central star of S~308. They also interpret this overabundance as an effect of the NeNa cycle. The simultaneous strong action of ON -- that we invoke to explain the O deficiency and almost normal C abundances seeing in NGC~6888 -- and NeNa cycles may be due to 
very high temperatures in the H-burning zones of the progenitor stars of our WR nebulae. It is interesting to note that one of the possible scenarios to explain the Na-O anticorrelation observed in globular clusters is based on the ejection of low-velocity material by previous generations of massive stars  \citep[e.g.][]{grattonetal12}. These stars should have experienced hot H-burning in order to alter the abundance of light elements with the simultaneous action of p-capture reactions of the CNO, NeNa, and MgAl chains \citep{denisenkovdenisenkova89}. Although the metallicities of the massive stars 
invoked to explain the pollution in globular clusters are extremely low in comparison with the solar one, it is very tempting to speculate about the possibility that objects similar to our WR ring nebulae may be the origin of the polluters of globular clusters. 

%%%%%%%%%%%%%%%%%%%%%%%%%%%%%%%%%%%%%%%%%%%%%%%%%%%%%%%%%%%%%%%%%%%%%%%%%%%%%%%%%%%%%%%%%%%%%%%%%
\section{Conclusions} \label{sec:conclusions}
%%%%%%%%%%%%%%%%%%%%%%%%%%%%%%%%%%%%%%%%%%%%%%%%%%%%%%%%%%%%%%%%%%%%%%%%%%%%%%%%%%%%%%%%%%%%%%%%%

We present very deep spectra of a sample of ring nebulae associated with evolved massive stars. Four of them associated to WR stars: NGC~6888, G2.4+1.4, RCW~58 and S~308 and two associated with O-type stars: NGC~7635 and RCW~52. The spectra have been obtained with OSIRIS and MagE spectrographs attached to the 10m GTC and the 6.5m Clay Telescope, respectively. We have obtained spectra for several slit positions 
or apertures in 
NGC~6888, NGC~7635 and G2.4+1.4 and for a single aperture in RCW~52, RCW~58 and S~308. We derive the physical conditions making use of several emission line intensity ratios as well as abundances for several ionic species from the intensity of CELs for all apertures. We obtain the first direct determination of {\elect} in the cases of RCW~52 and RCW~58. We derive the ionic abundances of C$^{2+}$ and/or O$^{2+}$ from faint RLs in several apertures of NGC~6888 and NGC~7635, permitting to derive their C/H and C/O ratios. This is the first time that {\oii} lines are observed in NGC~6888. The comparison of the O$^{2+}$/H$^{+}$ 
ratios calculated from CELs and RLs permit to estimate the ADF in NGC~6888 and NGC~7635, finding rather high values. The ADFs are larger than the typical ones of normal {\hii} regions in the Milky Way and other nearby spiral galaxies but similar to those found in the ionised gas of star-forming dwarf galaxies. 

We do not find compelling evidence of chemical inhomogeneities in the nebulae where we have 
observed several apertures: NGC~6888, NGC~7635 and G2.4+1.4. We obtain {\elect}({\foiii})= 19040$\pm$1130 K in a peripheral zone of NGC~6888, which is probably dominated by shock excitation. We find that since the two ring nebulae associated with O-type stars show normal chemical abundances, all the objects associated with WR stars show N enrichment. NGC~6888, RCW~58 and S~308 show also He enrichment and NGC~6888 and S~308 show some degree of O deficiency. The C abundance of NGC~6888 seems to be normal. This abundance pattern of the WR ring nebulae indicates the action of the nucleosynthesis of the CNO cycle and suggests an especially active ON cycle. NGC~6888 and G2.4+1.4 show lower Ne/O than expected, this may indicate the action of the NeNa cycle in a hot H-burning zone of the progenitor massive stars. 

We have compared abundance ratios of our four WR ring nebulae with the surface chemical composition predicted by stellar evolution models of massive stars with initial masses from 25 to 60 $M_{\odot}$ \citep{ekstrometal12, georgyetal12}. We find that non-rotational models of stars of initial masses between 25 and 40 $M_{\odot}$ seem to reproduce the abundance patterns in 
most of the objects, being this range wider in the case of S~308, from 25 to 50 $M_{\odot}$. Only rotational models of  25 $M_{\odot}$ show some fit with the data for NGC~6888, RCW~58 and S~308. As general result, the fit between models and  observational data is produced in a moment between the beginning of the SGR and before the onset of the WR phase. 

\section*{Acknowledgements}

This paper is based on observations made with the Gran Telescopio Canarias (GTC), installed in the Spanish Observatorio del Roque de los Muchachos of the Instituto de Astrof\'isica de Canarias, in the island of La Palma and in data gathered with the 6.5 meter Magellan Telescopes located at Las Campanas Observatory, Chile. We are grateful to Josef P\"ousel and Stefan Binnewies (www.capella-observatory.com), Daniel L\' opez (IAC/INT), Rick Gilbert (Star Shadows Remote Observatory, SSRO) and Don Goldman (www.astrodonimaging.com) for give us permission to reproduce the astronomical images used in figures~\ref{chart_2} and \ref{chart_4}. 
This work has been funded by the Spanish Ministerio de Econom\'\i a y Competitividad (MINECO) under project AYA2011-22614. JGR acknowledges support from Severo 
Ochoa excellence program (SEV-2011-0187) postdoctoral fellowship. AMD acknowledges support from the FONDECYT project 3140383 funded by the Chilean Government through CONICYT.

%%%%%%%%%%%%%%%%%%%%%%%%%%%%%%%%%%%%%%%%%%%%%%%%%%



%\bibliographystyle{mnras}
%\bibliography{cesar_bibliography} 

%%%%%%%%%%%%%%%%%%%%%%%%%%%%%%%%%%%%%%%%%%%%%%%%%%

% Don't change these lines
\bsp	% typesetting comment
\label{lastpage}
\end{document}